\documentclass[aps,prd,preprintnumbers,amsmath,amssymb,nofootinbib,11pt]{revtex4}
\usepackage{eepic}
\usepackage{indentfirst}
\usepackage{mathrsfs}
\usepackage{fancyhdr}
\usepackage{ulem}
\usepackage{float}
\usepackage{graphicx}
\usepackage[
colorlinks=true, linkcolor=black, breaklinks=true, urlcolor=blue,
citecolor=green]{hyperref}
\usepackage{epstopdf}
\usepackage{bm,bbm}

\usepackage{tabularx}
\usepackage{subfigure}
\usepackage{epsfig}
\usepackage{color}
\usepackage{slashed}
\usepackage{hyperref}

\newcommand{\be}{\begin{equation}}
\newcommand{\ee}{\end{equation}}
\renewcommand{\L}{\mathscr{L}}
\newcommand{\M}{\mathscr{M}}
\newcommand{\bra}{\langle}
\newcommand{\ket}{\rangle}
\newcommand{\nn}{\nonumber}
\newcommand{\MeV}{\,\text{MeV}}
\newcommand{\GeV}{\,\text{GeV}}
\renewcommand{\vec}[1]{\mathbf{#1}}
\newcommand{\diff }{{\text{d}}}
\newcommand{\qperp}{\vec q_\perp}
\newcommand{\q}{\vec q}

\begin{document}
\pagestyle{plain}

\title {\boldmath Effects of $Z_b$ states and bottom meson loops on\\ $\Upsilon(4S) \to \Upsilon(1S,2S) \pi^+\pi^-$ transitions}

\author{ Yun-Hua~Chen$^a$}
\author{ Martin Cleven$^b$}
\author{ Johanna T.\ Daub$^{a}$}
\author{ Feng-Kun Guo$^{c,d}$}
\author{ Christoph Hanhart$^{e}$}
\author{ Bastian Kubis$^{a}$}
\author{ Ulf-G.\ Mei\ss ner$^{a,e}$}
\author{ Bing-Song~Zou$^{c,d}$}
\affiliation{${}^a$Helmholtz-Institut f\"ur Strahlen- und Kernphysik (Theorie) and\\
             Bethe Center for Theoretical Physics,
             Universit\"at Bonn,
             53115 Bonn, Germany\\
              ${}^b$Departament de Fisica Quantica i Astrofisica
                    Universitat de Barcelona, 08028-Barcelona, Spain\\
             ${}^c$CAS Key Laboratory of Theoretical Physics,
             Institute of Theoretical Physics,\\ Chinese Academy of Sciences,
Beijing 100190, China \\
             ${}^d$School of Physical Sciences, University of Chinese Academy of
             Sciences, Beijing 100049, China\\
             ${}^e$Institut f\"ur Kernphysik, Institute for Advanced
Simulation, and\\
             J\"ulich Center for Hadron Physics,
             Forschungszentrum J\"ulich,
             52425 J\"ulich, Germany
}

\begin{abstract}

We study the dipion transitions $\Upsilon(4S) \rightarrow
\Upsilon(nS) \pi^+\pi^-$ $(n=1,2)$. In particular, we consider the
effects of the two intermediate bottomoniumlike exotic states
$Z_b(10610)$ and $Z_b(10650)$ as well as bottom meson loops. The
strong pion--pion final-state interactions, especially including
channel coupling to $K\bar{K}$ in the $S$-wave, are taken into
account model-independently by using dispersion theory. Based on a
nonrelativistic effective field theory we find that the contribution
from the bottom meson loops is comparable to those from the chiral
contact terms and the $Z_b$-exchange terms. For the $\Upsilon(4S)
\rightarrow \Upsilon(2S) \pi^+\pi^-$ decay, the result shows that
including the effects of the $Z_b$-exchange and the bottom meson
loops can naturally reproduce the two-hump behavior of the $\pi\pi$
mass spectra. Future angular distribution data are decisive for the
identification of different production mechanisms. For the
$\Upsilon(4S) \rightarrow \Upsilon(1S) \pi^+\pi^-$ decay, we show
that there is a narrow dip around $1\GeV$ in the $\pi\pi$ invariant
mass distribution, caused by the final-state interactions. The
distribution is clearly different from that in similar transitions
from lower $\Upsilon$ states, and needs to be verified by future
data with high statistics. Also we predict the decay width and the
dikaon mass distribution of the $\Upsilon(4S) \rightarrow
\Upsilon(1S) K^+ K^-$ process.

\end{abstract}

\maketitle

\newpage
\section{Introduction}
The processes of dipion emission of the bottomonia $\Upsilon(mS) \to
\Upsilon(nS) \pi \pi $ are important for understanding the
heavy-quarkonium dynamics and low-energy QCD. Because the bottomonia
are expected to be nonrelativistic and compact, the method of the
QCD multipole
expansion~\cite{Voloshin1980,Novikov1981,Kuang1981,Kuang2006} is
often used to study these transitions, where the pions emitted come
from the hadronization of soft gluons. Though successful in
describing many $\Upsilon(mS) \to \Upsilon(nS) \pi \pi $ processes,
a well-known anomaly about the method of the QCD multipole expansion
is that it cannot reproduce the two-hump behavior in the
experimental $\pi\pi$ invariant mass spectra of the decays
$\Upsilon(3S) \rightarrow \Upsilon(1S) \pi\pi$ and $\Upsilon(4S)
\rightarrow \Upsilon(2S)
\pi^+\pi^-$~\cite{Eichten2008,Guo2005,Guo2007:4S}. In a previous
study~\cite{Chen2016}, we found that by including the effects of the
two bottomoniumlike exotic states $Z_b(10610)$ and $Z_b(10650)$
discovered by the Belle Collaboration~\cite{Belle2011:1,Belle2012:1}
as well as the $\pi\pi$ final-state interaction (FSI), the anomaly
of the $\Upsilon(3S) \rightarrow \Upsilon(1S) \pi\pi$ process can be
naturally explained. Such an analysis is a modern version of the
much earlier studies in
Refs.~\cite{Voloshin:1982ij,Anisovich:1995zu}, where an isovector
$b\bar b q\bar q$ state was considered. Although the direct decay of
$Z_b$ into $\Upsilon(4S)\pi$ is kinematically impossible, it may be
illuminating to analyze the effect of the $Z_b$-exchange mechanism
in the $\Upsilon(4S) \rightarrow \Upsilon(1S,2S) \pi^+\pi^-$
processes, which is performed in this study. In this context it is
important to note that improved data on $\Upsilon(nS)$ decays are to
be expected from Belle-II that will start operation soon---for a
detailed discussion of prospects for various decays relevant
for this study we refer to Ref.~\cite{Bondar:2016hva}.

The $\Upsilon(4S)$ meson is above the $B\bar{B}$ threshold and
decays predominantly to $B\bar{B}$, so loop effects with
intermediate bottom mesons may play an important role in
$\Upsilon(4S) \to \Upsilon(nS) \pi\pi$  $(n=1, 2)$. Also, the
inclusion of the loops will introduce non-analyticities arising from
the $B\bar B$ threshold needed to be taken into account in
dispersion theory, which will be discussed later.  Because the
bottomonia are close to the open-bottom meson production threshold,
the velocity of the intermediate bottom mesons is small and can be
treated as an expansion parameter to build power-counting rules in a
nonrelativistic effective field theory
(NREFT)~\cite{Guo2009:PRL,Guo2011:effect,Martin2013}. Within the
NREFT scheme, we will calculate the dominant box diagrams in the
dipion emissions of $\Upsilon(4S)$, and find that their contribution
is comparable in size to the chiral contact terms and the
$Z_b$-exchange graphs.

In the $\Upsilon(4S) \to \Upsilon(1S) \pi\pi$ process, the dipion invariant mass
reaches above the $K\bar{K}$ threshold, so the coupled-channel FSI in the
$S$-wave is strong and needs to be taken into account. Based on analyticity and
unitarity, dispersion theory can achieve this in a model-independent way. In
this study, we will use dispersion theory in the form of modified Omn\`es
solutions, in which the left-hand-cut contribution is approximated by the sum of
the $Z_b$-exchange mechanism and the bottom meson loops.
At low energies, the amplitude should agree with the leading chiral results, so
the subtraction functions can be determined by matching to chiral contact terms.
For the leading contact couplings of two $S$-wave bottomonia to an even number
of light pseudoscalar mesons, we will adopt the Lagrangian given in
Ref.~\cite{Mannel}, constructed in the spirit of the chiral and the heavy-quark
nonrelativistic expansions.

This paper is organized as follows. In Sec.~\ref{theor}, we present
the theoretical framework and elaborate on the calculation of the
amplitudes as well as the dispersive treatment of the FSI. In
Sec.~\ref{pheno}, we fit the experimental data of the $\pi\pi$
invariant mass distribution to determine the coupling constants, and
discuss the contributions of different mechanisms. A summary will be
given in Sec.~\ref{conclu}.
\section{Theoretical framework}\label{theor}
\subsection{Lagrangians}
Because in the heavy-quark limit the spin of the heavy quarks
decouples, it is convenient to introduce the heavy quarkonia and
heavy hadrons in terms of spin multiplets. One has $J \equiv
\vec{\Upsilon} \cdot \boldsymbol{\sigma}+\eta_b$, where
$\vec\Upsilon$ and $\eta_b$ annihilate the $\Upsilon$ and $\eta_b$
states, respectively, and $\boldsymbol{\sigma}$ contains the Pauli
matrices~\cite{Guo2011}. The bottom mesons are collected in
$H_a=\vec{V}_a \cdot \boldsymbol{\sigma}+P_a$ with
$P_a(V_a)=(B^{(*)-},\bar{B}^{(*)0},\bar{B}_s^{(*)0})$, and
$\bar{H}_a=- \bar{\vec{V}}_a \cdot \boldsymbol{\sigma}+\bar{P}_a$
with
$\bar{P}_a(\bar{V}_a)=(B^{(*)+},B^{(*)0},B_s^{(*)0})$~\cite{Mehen2008}.

The effective Lagrangian for the contact
$\Upsilon\Upsilon^{\prime}\pi\pi$ and
$\Upsilon\Upsilon^{\prime}K\bar{K}$ coupling, at the lowest order
in the chiral as well as the heavy-quark expansion,
reads~\cite{Mannel,Chen2016}
\begin{equation}\label{LagrangianUpUppipi}
\L_{\Upsilon\Upsilon^{\prime}\Phi\Phi} = \frac{c_1}{2}\bra J^\dagger
J^\prime \ket \bra u_\mu u^\mu\ket +\frac{c_2}{2}\bra J^\dagger
J^\prime \ket \bra u_\mu u_\nu\ket v^\mu v^\nu +\mathrm{h.c.} \,,
\end{equation}
where $v^\mu=(1,\vec{0})$ is the velocity of the heavy quark. The
Goldstone bosons of the spontaneous breaking of chiral symmetry can
be parametrized as
\begin{align}
u_\mu &= i \left( u^\dagger\partial_\mu u\, -\, u \partial_\mu u^\dagger\right) \,, \qquad
u = \exp \Big( \frac{i\Phi}{\sqrt{2}F} \Big)\,, \nonumber\\
\Phi &=
 \begin{pmatrix}
   {\frac{1}{\sqrt{2}}\pi ^0 +\frac{1}{\sqrt{6}}\eta _8 } & {\pi^+ } & {K^+ }  \\
   {\pi^- } & {-\frac{1}{\sqrt{2}}\pi ^0 +\frac{1}{\sqrt{6}}\eta _8} & {K^0 }  \\
   { K^-} & {\bar{K}^0 } & {-\frac{2}{\sqrt{6}}\eta_8 }  \\
 \end{pmatrix} , \label{eq:u-phi-def}
\end{align}
where $F$ is the pseudo-Goldstone boson decay constant, and we will
use $F_\pi=92.2\MeV$ for the pions and $F_K=113.0\MeV$ for the
kaons. The two operators in Eq.~\eqref{LagrangianUpUppipi} both
scale as $\mathcal{O}(q_\pi^2)$ in the expansion in (soft) pion
momenta $q_\pi$.

The leading Lagrangian for the  $Z_b \Upsilon\pi$ interaction, which
is needed in the calculation of the mechanism $\Upsilon(mS) \to
Z_b\pi \to\Upsilon(nS) \pi\pi$, reads~\cite{Guo2011}\footnote{Here
we only include the terms relevant to the $\Upsilon$ coupling rather
than the full spin multiplet defined before as
$J=\vec{\Upsilon}\cdot\vec{\sigma}+\eta_b$. In this way, we avoid
the discussion of the internal spin structure of the $Z_b$ states,
which depends on specific models for $Z_b$ and is not really settled
yet.}
\be\label{LagrangianZbUppi} \L_{Z_b\Upsilon\pi} =
\sum_{j=1,2}\sum_{l}C_{Z_{bj}\Upsilon(lS)\pi} \Upsilon^i(lS) \bra
{Z^i_{bj}}^\dagger u_\mu \ket v^\mu +\mathrm{h.c.} \,,
\ee
where
$Z_{b1}$ and $Z_{b2}$ are used to refer to $Z_b(10610)$ and
$Z_b(10650)$, respectively. The $Z_b$ states are collected in the
matrix as
\begin{equation} \label{eq:Z-field}
Z^i_{bj}=
 \left( {\begin{array}{*{3}c}
   \frac{1}{\sqrt{2}}Z^{0i}_{bj} & Z^{+i}_{bj} & 0  \\
   Z^{-i}_{bj} & -\frac{1}{\sqrt{2}}Z^{0i}_{bj} & 0 \\
   0 & 0 & 0
\end{array}} \right)\,.
\end{equation}
Note that since strange partners of the $Z_b$ states, $Z_{bs}$, have
not been observed, we set the corresponding matrix entries in
Eq.~\eqref{eq:Z-field} to zero.

To calculate the box diagrams, we need the Lagrangian for the
coupling of the $S$-wave bottomonium fields to the bottom and
antibottom mesons~\cite{Guo2009:PRL},
\begin{equation}\label{LagrangianJHH}
\L_{JHH}=\frac{i\, g_{JHH}}{2}\bra J^\dag H_a
\boldsymbol{\sigma}\cdot \!\overleftrightarrow{\partial}\!
\bar{H}_a\ket + {\rm h.c.}\,,
\end{equation}
where $A\overleftrightarrow{\partial}\!B\equiv
A(\overrightarrow{\partial}B)-(\overrightarrow{\partial}A)B$. We
also need the Lagrangian for the axial coupling of the Goldstone
bosons to the bottom and antibottom mesons, which at leading order
in heavy-flavor chiral perturbation theory is given
by~\cite{Burdman:1992gh,Wise:1992hn,Yan:1992gz,Casalbuoni:1996pg,Mehen2008}
\begin{equation}\label{LagrangianHHPhi}
\L_{HH\Phi}= \frac{g_\pi}{2} \bra \bar{H}_a^\dagger  \boldsymbol{\sigma} \cdot \vec{u}_{ab}
\bar{H}_b\ket -\frac{g_\pi}{2} \bra H_a^\dagger H_b
\boldsymbol{\sigma} \cdot \vec{u}_{ba} \ket,
\end{equation}
where $u^i=-\sqrt{2}\partial^i \Phi/F + \mathcal{O}(\Phi^3)$ corresponds to the three-vector components of $u_\mu$
as defined in Eq.~\eqref{eq:u-phi-def}.
Here we will use $g_\pi=0.5$ from a recent lattice QCD
calculation~\cite{Bernardoni:2014kla}.\footnote{The precise value quoted in
Ref.~\cite{Bernardoni:2014kla} is $g_\pi=0.492\pm0.029$.}

\subsection{Power counting of the loops}

Since the $\Upsilon(4S)$ is above the $B\bar{B}$ threshold and
decays predominantly into $B\bar{B}$ pairs, the loop mechanism with
intermediate bottom mesons may play a significant role in the
bottomonium transitions $\Upsilon(4S) \rightarrow \Upsilon(nS)
\pi^+\pi^-$. In this section, we will analyze the power counting of
different kinds of loops, based on
NREFT~\cite{Guo2009:PRL,Guo2011:effect,Martin2013}. In NREFT, the
expansion parameter is the typical velocity of the intermediate
heavy meson, namely
$\nu=\sqrt{|m_{\Upsilon(lS)}-m_{B^{(*)}}-m_{B^{(*)}}|/m_{B^{(*)}}}$,
and $\nu\ll 1$  since  $\Upsilon(lS)$ are close to the
$B^{(*)}\bar{B}^{(*)}$ thresholds. Each nonrelativistic propagator
is counted as $1/\nu^2$, and the full integral measure $\int d^4 l $
as $\nu^5$. More details of the power counting rules are
elaborated in Ref.~\cite{Guo2011:effect}.

\begin{figure}
\centering
\includegraphics[width=\linewidth]{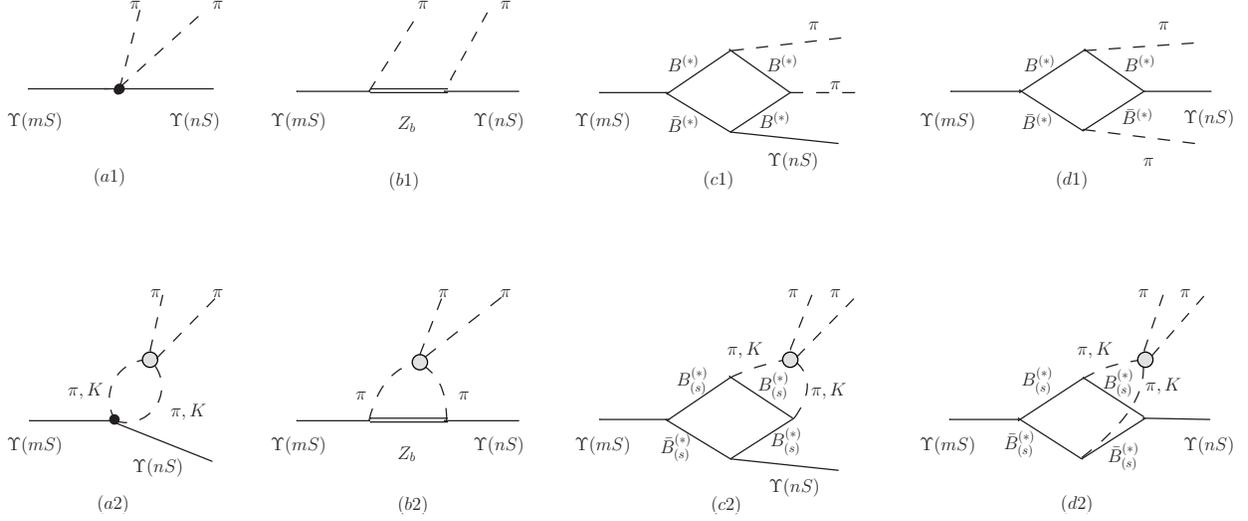}
\caption{Feynman diagrams considered for the $\Upsilon(mS)
\rightarrow \Upsilon(nS) \pi \pi $ processes. The crossed diagrams
of (b1), (c1), (b2), and (c2) are not shown explicitly. The gray
blob denotes the final-state interaction.
}\label{fig.FeynmanDiagram}
\end{figure}
\begin{figure}
\centering
\includegraphics[width=\linewidth]{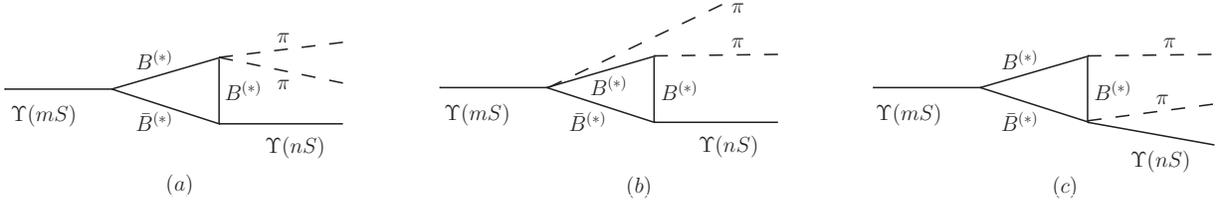}
\caption{Subleading contributions for $\Upsilon(mS)
\rightarrow \Upsilon(nS) \pi \pi $ that are suppressed in comparison to the four-point functions
in Fig.~\ref{fig.FeynmanDiagram} and hence not considered in the calculations.
The corresponding power counting arguments are given in the main text.}
\label{fig.FeynmanDiagramTriangle}
\end{figure}

Without considering the FSI, there are five different kinds of loop
contributions, namely the box diagrams displayed in
Fig.~\ref{fig.FeynmanDiagram}\,(c1), (d1), and the triangle diagrams
displayed in Fig.~\ref{fig.FeynmanDiagramTriangle}\,(a)--(c). We analyze them
one by one as follows:
\begin{enumerate}
\item
Box diagrams, namely Fig.~\ref{fig.FeynmanDiagram}\,(c1), (d1): As
indicated in Eq.~\eqref{LagrangianHHPhi}, the vertex for the axial
coupling of the pion to the bottom mesons is proportional to the
external momentum of the pion $q_\pi$. Both the vertices for the
initial and final bottomonia are in a $P$-wave, and the product of the
two vertices can be counted as $\mathcal{O}(\nu^2)$, so the box
diagrams are counted as $\nu^5 \nu^2 q_\pi^2/\nu^8=q_\pi^2/\nu.$
Note that these contributions thus have the same scaling in pion momenta as
the leading $\Upsilon\Upsilon^{\prime}\pi\pi$ contact terms from the Lagrangian Eq.~\eqref{LagrangianUpUppipi},
but are formally enhanced by $1/\nu$ in the non-relativistic velocity parameter.

\item
Fig.~\ref{fig.FeynmanDiagramTriangle}\,(a): The leading
$B^{(*)}B^{(*)}\pi\pi$ vertex comes from the covariant chiral
derivative term $\bra H_a^\dag(iD_0)_{ba}H_b\ket =\bra
H_a^\dag(i\partial_0-iV_0)_{ba}H_b\ket$~\cite{Stewart,Mehen2006},
in which the pion pair produced by the vector current,
$V^\mu=\frac{1}{2}(u^\dag \partial^\mu u+u\partial^\mu u^\dag)$, cannot form a
positive-parity and $C$-parity state, so this leading vertex does not contribute to the
$\Upsilon(mS) \rightarrow \Upsilon(nS) \pi \pi $ processes.
Isoscalar, $PC=++$ pion pairs only enter in the next order
$\mathcal{O}(q_\pi^2)$ from point vertices. For the vertices of the
initial and final bottomonia, both of them are in $P$-waves, so the
product of them can be counted as $\mathcal{O}(\nu^2)$. These diagrams
hence count as $\nu^5 \nu^2 q_\pi^2/\nu^6=\nu q_\pi^2$, and are suppressed
compared to the contact terms $\propto c_{1,2}$ by the factor $\nu$.

\item
Fig.~\ref{fig.FeynmanDiagramTriangle}\,(b), (c): The leading
$\Upsilon(lS)B^{(*)}\bar{B}^{(*)}\pi$ vertex given by $\bra J \bar{H}_a^\dag
H_b^\dag\ket u_{ab}^0$~\cite{Mehen2013}
is proportional to the energy of the pion, $E_\pi\sim q_\pi$. In
Fig.~\ref{fig.FeynmanDiagramTriangle}\,(b), the vertex for the
initial bottomonium is in an $S$-wave, and the vertex for the final
bottomonium is in a $P$-wave, so the loop momentum must contract with
the external momentum $q_\pi$ and hence the $P$-wave vertex scales
as $\mathcal{O}(q_\pi)$. For this reason,
Fig.~\ref{fig.FeynmanDiagramTriangle}\,(b) is counted as
$\nu^5 q_\pi^3/(\nu^6 m_B)=q_\pi^3/(\nu m_B)$, where the factor $m_B$ has been
introduced to match the dimension with the scaling for cases~1 and~2.
Analogous arguments hold for Fig.~\ref{fig.FeynmanDiagramTriangle}\,(c).
This class of diagrams is therefore suppressed in the chiral expansion in pion momenta,
compared to the $c_{1,2}$ terms.
\end{enumerate}
We find thus that according to the power counting the box diagrams are dominant
among the loop contributions, and the only ones not expected to be suppressed relative
to the tree-level contact terms.
We will therefore only calculate these in the present study.
Note that all (box and triangle) loop contributions discussed here are ultraviolet-finite,
and do not require the additional introduction of counterterms.

\subsection{Tree-level amplitudes and box diagram calculation}

First we define the Mandelstam variables in the decay process of $
\Upsilon(mS)(p_a) \to \Upsilon(nS)(p_b) P(p_c)P(p_d) $
\begin{align}
s &= (p_c+p_d)^2 , \qquad
t_P=(p_a-p_c)^2\,, \qquad u_P=(p_a-p_d)^2\,,\nn\\
3s_{0P}&\equiv s+t_P+u_P=
 m_{\Upsilon(mS)}^2+m_{\Upsilon(nS)}^2+2m_P^2  \,,
\end{align}
where $P$ denotes the pseudoscalar $\pi$ or $K$, since we also need
to take into account the virtual process $ \Upsilon(mS)(p_a) \to
\Upsilon(nS)(p_b) K(p_c) \bar K(p_d) $ in the coupled-channel FSI. $t_P$
and $u_P$ can be expressed in terms of $s$ and the helicity angle
$\theta$ according to
\begin{align}
t_P &= \frac{1}{2} \left[3s_{0P}-s+\kappa_P(s)\cos\theta \right]\,,&
u_P &= \frac{1}{2} \left[3s_{0P}-s-\kappa_P(s)\cos\theta \right]\,, \nn\\
\kappa_P(s) &\equiv \sigma_P
\lambda^{1/2}\big(m_{\Upsilon(mS)}^2,m_{\Upsilon(nS)}^2,s\big) \,, &
\sigma_P &\equiv \sqrt{1-\frac{4m_P^2}{s}} \,, \label{eq:tu}
\end{align}
where $\theta$ is defined as the angle between the initial
$\Upsilon(mS)$ and the positive pseudoscalar in the rest frame of
the $PP$ system, and $\lambda(a,b,c)=a^2+b^2+c^2-2(ab+ac+bc)$. We
define $\vec{q}$ as the 3-momentum of the final bottomonium in the
rest frame of the initial state with
\be \label{eq:q}
|\vec{q}|=\frac{1}{2m_{\Upsilon(mS)}}
\lambda^{1/2}\big(m_{\Upsilon(mS)}^2,m_{\Upsilon(nS)}^2,s\big) \,.
\ee

The calculation of the tree amplitudes is very similar to our
previous study of $\Upsilon(3S)$ decays~\cite{Chen2016}, so here we
just quote the partial-wave-projected results. Parity and $C$-parity
conservation require the pion pair to have even relative angular
momentum $l$. We will only consider the $S$-wave and $D$-wave
components in this study, neglecting the effects of higher partial
waves. For the $S$-wave, the amplitudes of the chiral contact term
and the $Z_b$-exchange term read
\begin{align}\label{eq.M0chiral}
M_0^{\chi,P}(s)&=-\frac{2}{F_P^2}\sqrt{m_{\Upsilon(mS)}m_{\Upsilon(nS)}}
\bigg\{c_1 \left(s-2m_P^2 \right)
+\frac{c_2}{2} \bigg[s+\vec{q}^2\Big(1  -\frac{\sigma_P^2}{3}
\Big)\bigg]\bigg\}\,, \\
\hat{M}_0^{Zb,\pi}(s)&=-\frac{2
\sqrt{m_{\Upsilon(mS)}m_{\Upsilon(nS)}}}{F_\pi^2
\kappa_\pi(s)}\sum_{i=1,2}m_{Z_{bi}}C_{mn,i}\Big\{
\big(s+|\vec{q}|^2\big)Q_0(y_{\pi i})-|\vec{q}|^2\sigma_\pi^2\big[y_{\pi i}^2
Q_0(y_{\pi i})-y_{\pi i}\big] \Big\}\,,\label{eq.M0hat}
\end{align}
where $C_{mn,i}\equiv C_{Z_{bi}\Upsilon(mS)\pi}
C_{Z_{bi}\Upsilon(nS)\pi}$, $y_{\pi i} \equiv
{(3s_{0\pi}-s-2m_{Z_{bi}}^2)}/{\kappa_\pi(s)}$, and $Q_0(y)$ is a Legendre
function of the second kind,
\begin{equation}
Q_0(y)=\frac{1}{2}\int_{-1}^1 \frac{\diff z}{y-z}P_0(z)
 = \frac{1}{2}\log \frac{y+1}{y-1}
\end{equation}
($P_i(z)$ refers to the standard Legendre polynomials).
Note again that we consider the $Z_b$-exchange diagrams only for the process involving
pions. For every heavy particle, namely the bottomonia and the $Z_b$
states here, a nonrelativistic normalization factor $\sqrt{M}$ has
been multiplied to the expressions, with $M$ the corresponding mass.
The widths of the $Z_b$ states are neglected in the present
calculation, since their nominal values are of the order of $10\MeV$ and thus much
smaller than the gap between their masses and the $\Upsilon(lS)\pi$ threshold.

For the $D$-wave, in which $\pi\pi$ scattering is elastic to very good approximation in the energy range
considered, we only consider the single-channel FSI, and therefore we
just give the amplitudes of the process involving pions,
\begin{align}\label{eq.M2chiral}
M_2^{\chi,\pi}(s)&=\frac{2}{3
F_\pi^2}\sqrt{m_{\Upsilon(mS)}m_{\Upsilon(nS)}}\,c_2
|\vec{q}|^2\sigma_\pi^2\,,
\\
\hat{M}_2^{Zb,\pi}(s)&=-\frac{5\sqrt{m_{\Upsilon(mS)}m_{\Upsilon(nS)}}}{F_\pi^2\kappa_\pi(s)}
\sum_{i=1,2}m_{Z_{bi}}C_{mn,i}
\big[s+|\vec{q}|^2-|\vec{q}|^2\sigma_\pi^2 y_{\pi i}^2\big]
\big[(3y_{\pi i}^2-1)Q_0(y_{\pi i})-3y_{\pi i}\big]\,.\label{eq.M2hat}
\end{align}

\begin{table*}
\begin{center}
\renewcommand{\arraystretch}{1.3}
\begin{tabular}{l|l}\hline\hline
Type 1(c1)& $[B,\bar{B},B,B^*],[B,\bar{B}^*,B,B^*],[B^*,\bar{B},B^*,B],[B,\bar{B}^*,B^*,B^*],$\\
          & $[B^*,\bar{B},B^*,B^*],[B^*,\bar{B}^*,B,B^*],[B^*,\bar{B}^*,B^*,B],[B^*,\bar{B}^*,B^*,B^*]$\\
Type 1(d1)& $[B,\bar{B},\bar{B}^*,B^*],[B,\bar{B}^*,\bar{B},B^*],[B^*,\bar{B}^*,\bar{B},B],[B^*,\bar{B},\bar{B}^*,B], [B,\bar{B}^*,\bar{B}^*,B^*],$\\
          & $[B^*,\bar{B},\bar{B}^*,B^*],[B^*,\bar{B}^*,\bar{B},B^*], [B^*,\bar{B}^*,\bar{B}^*,B],[B^*,\bar{B}^*,\bar{B}^*,B^*]$.\\
\hline\hline%
\end{tabular}
\caption{\label{tab:loops}All loops contributing in each diagram class. The mesons are listed as $[M1,M2,M3,M4]$, type 1(c1) and type 1(d1) refer to the corresponding diagrams in Fig.~\ref{fig.FeynmanDiagram}.
Two more configurations appear as type 1(c1) in principle,
namely $[B,\bar{B},B^*,B^*]$ and $[B^*,\bar{B},B,B^*]$, however, their contributions to amplitude $\M_1$,
see Eq.~\eqref{eq:Mdecomp}, vanishes, and hence they are strongly suppressed.
Flavor labels are dropped for simplicity.}
\end{center}
\end{table*}

Now we briefly discuss the calculation of the box diagrams. There
are four intermediate bottom mesons in the box diagrams
Fig.~\ref{fig.FeynmanDiagram}\,(c1) and (d1), where we denote the
top left one as $M1$, and the others as $M2$, $M3$, and $M4$, in
counterclockwise order. The individual contributions are listed in
Table~\ref{tab:loops}, with the pseudoscalar or vector content of
$[M1,M2,M3,M4]$ explicitly shown. For the $\Upsilon(mS) \rightarrow
\Upsilon(nS) K \bar{K} $ processes, some intermediate states can be
strange bottom mesons $B_s^{(*)}$, and there are four possibilities
for each $[M1,M2,M3,M4]$ given above. For simplicity, we do not list
the combinations of intermediate states in the $\Upsilon(mS)
\rightarrow \Upsilon(nS) K \bar{K} $ processes explicitly.

The general amplitude for the process $ \Upsilon(mS)(p_a) \to
\Upsilon(nS)(p_b) P(p_c)P(p_d) $ reads \be \M(\Upsilon(mS) \to
\Upsilon(nS) P P)=\epsilon_{\Upsilon(mS)}^i
\epsilon_{\Upsilon(nS)}^j \M^{ij}\big(\Upsilon(mS) \to \Upsilon(nS)
PP \big)\,, \ee and $\M^{ij}(\Upsilon(mS) \to \Upsilon(nS) P P)$ can
be decomposed as \be \label{eq:Mdecomp} \M^{ij}(\Upsilon(mS) \to
\Upsilon(nS) P P) = \delta^{ij}\M_1+ \ldots \,, \ee where we have
omitted the remaining terms proportional to tensor structures built
from the different momenta. For the loop amplitude, we have checked
that the $\M_1$ term is indeed numerically dominant, which agrees
with the argument that other contractions of the polarization
vectors are suppressed in the heavy-quark nonrelativistic expansion.
So in the following we will only keep the terms proportional to
$\bm{\epsilon}_{\Upsilon(mS)}\cdot \bm{\epsilon}_{\Upsilon(nS)}$, as
we did for the tree amplitude. Details on the analytic calculation
of the box diagrams are given in Appendix~\ref{app:box}.

\begin{figure}
\centering
\includegraphics[width=\linewidth]{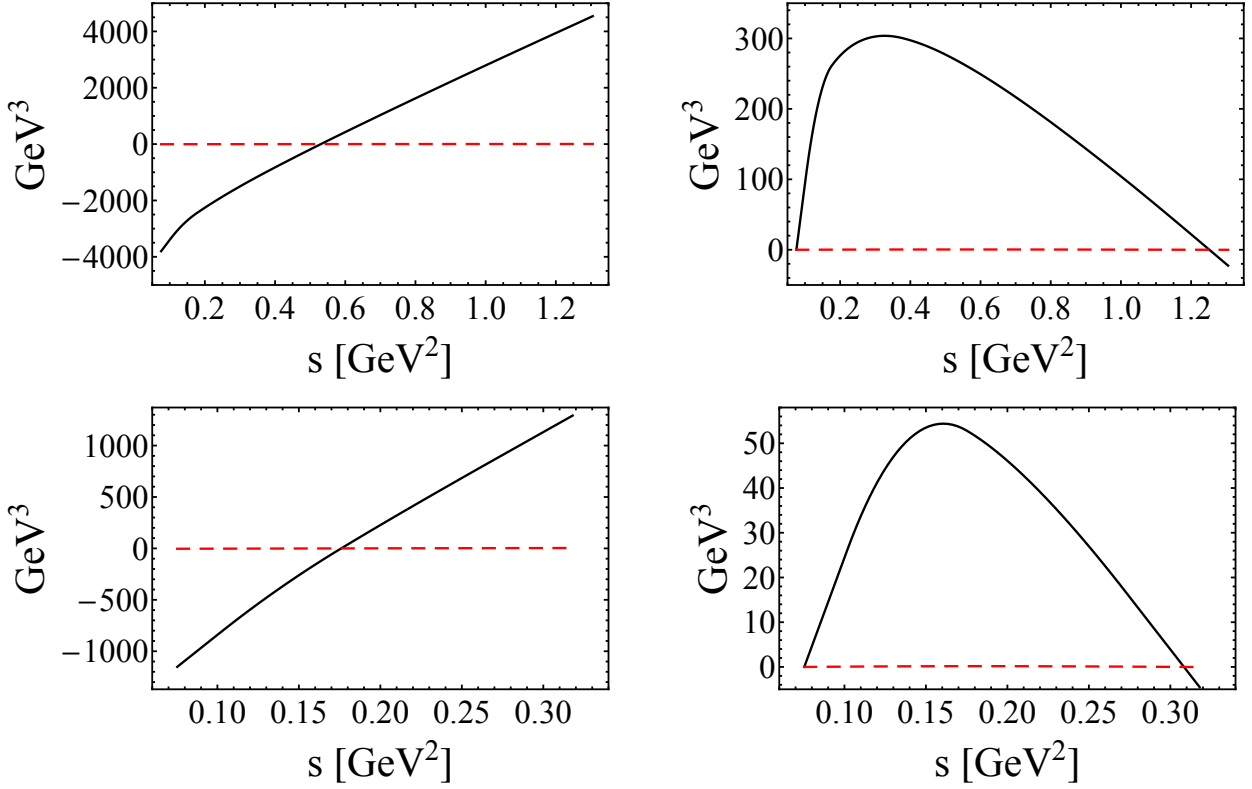}
\caption{The $S$- (left) and $D$-wave (right) box amplitudes
in $\Upsilon(4S) \to \Upsilon(1S) \pi^+\pi^-$ (top) and
$\Upsilon(4S) \to \Upsilon(2S) \pi^+\pi^-$ (bottom). The black
solid and red dashed lines denote the real and imaginary parts,
respectively. }\label{fig.BoxLoopAmplitude4S}
\end{figure}

The partial-wave projection of the loop amplitude for the  $
\Upsilon(mS) \to \Upsilon(nS) P P $ process can be denoted as
\begin{equation}
\hat{M}_{l}^{\text{loop},P}(s)=g_{JHH(mS)}\,g_{JHH(nS)}\, \text{AmpBox}_{l}^P(s)\,.
\end{equation}
The analytic expressions of
$\text{AmpBox}_{l}^P(s)$ are very involved, so in
Fig.~\ref{fig.BoxLoopAmplitude4S} we only
plot the numerical results for $\Upsilon(4S) \rightarrow
\Upsilon(1S,2S) \pi^+\pi^-$ in the physical region.
Note that the imaginary parts, which are due to the on-shell $B\bar{B}$ intermediate states,
are tiny due to the smallness of phase space and the fact that the $B\bar{B}$ pair appears
in a relative $P$-wave.

\subsection{Final-state interactions with a dispersive approach, Omn\`es solution }

There are strong FSIs in the $\pi\pi$ system especially in the
isoscalar $S$-wave, which can be taken into account
model-independently using dispersion theory. Based on the principles
of unitarity and analyticity, dispersion theory determines the decay
amplitudes up to some subtraction constants, which can be fixed by
matching to the results of chiral effective theory. Since the mass
difference between the $\Upsilon(4S)$ and the $\Upsilon(1S)$ is larger than the
$K\bar{K}$ threshold, we will consider the isospin symmetric two-channel
($\pi\pi$ and $K\bar K$) FSI for the dominant $S$-wave component, while for the
$D$-wave only single-channel $\pi\pi$ FSI will be considered.

For the $\Upsilon(mS) \rightarrow \Upsilon(nS) \pi^+ \pi^-$
processes, the partial-wave expansion of the amplitude including FSI
reads \be \M^\text{full}(s,\cos\theta) =\epsilon_{\Upsilon(nS)}\cdot
\epsilon_{\Upsilon(mS)} \sum_{l=0}^{\infty}
\left[M_l^\pi(s)+\hat{M}_l^\pi(s)\right] P_l(\cos\theta)\,, \ee
where $M_l^\pi(s)$ represents the right-hand cut part and accounts
for $s$-channel rescattering, and the ``hat functions''
$\hat{M}_l^\pi(s)$ contain the left-hand cuts, contributed by
crossed-channel pole terms or open-flavor loop effects. In general
the box diagrams contribute to both the left-hand cuts at $t, u
> (m_{B^{(*)}}+m_{B^{(*)}})^2$ and right-hand cut at $s >
(m_{B^{(*)}}+m_{B^{(*)}})^2$, however, this right-hand cut is far
away from the physical region, so it can be safely neglected. In
this study, we approximate the left-hand cuts by the sum of the
$Z_b$-exchange diagram and the box diagrams,
$\hat{M}_l^\pi(s)=\hat{M}_l^{Z_b,\pi}(s)+\hat{M}_l^{\text{loop},\pi}(s)$.
Similar methods to approximate the left-hand-cut structures by
including resonance exchange (in the case of no loops) have been
applied in Refs.~\cite{Moussallam-gamma,KubisPlenter,ZHGuo,Kang}.

Next we discuss the Omn\`es solution to obtain the amplitude
including FSI. For simplicity first we discuss the single-channel
solution, which applies for the $D$-wave case. The functions
$\hat{M}_l(s)$ are real along the right-hand cut, so in the elastic
$\pi\pi$ rescattering region the partial-wave unitarity conditions
reads
\begin{equation}\label{eq.unitarity1channel}
\textrm{Im}\, M_l(s)= \left[M_l(s)+\hat{M}_l(s)\right]
\sin\delta_l^0(s) e^{-i\delta_l^0(s)}\,.
\end{equation}
In the elastic region, the phases $\delta_l^I$ of the partial-wave amplitudes
of isospin $I$ and angular momentum $l$ equal the
$\pi\pi$ elastic phase shifts modulo $n\pi$, as required by Watson's
theorem~\cite{Watson1,Watson2}. The Omn\`es function is defined
as~\cite{Omnes}
\begin{equation}\label{Omnesrepresentation}
\Omega_l^I(s)=\exp
\bigg\{\frac{s}{\pi}\int^\infty_{4m_\pi^2}\frac{\diff x}{x}
\frac{\delta_l^I(x)}{x-s}\bigg\}\,,
\end{equation}
which obeys
$\Omega_l^I(s+i\epsilon)=e^{2i\delta_l^I}\Omega_l^I(s-i\epsilon)$.
Using the Omn\`es function, the solution of
Eq.~\eqref{eq.unitarity1channel} can be
obtained~\cite{Leutwyler96,Chen2016}
\be\label{OmnesSolution1channel}
M_l(s)=\Omega_l^0(s)\bigg\{P_l^{n-1}(s)+\frac{s^n}{\pi}\int_{4m_\pi^2}^\infty
\frac{\diff x}{x^n} \frac{\hat
M_l(x)\sin\delta_l^0(x)}{|\Omega_l^0(x)|(x-s)}\bigg\} \,, \ee where
the polynomial $P_l^{n-1}(s)$ is a subtraction function. For the
$D$-wave phase shift $\delta_2^0(s)$, we will use the result given
by the Madrid--Krak\'ow collaboration~\cite{Pelaez}, and smoothly
continue it to $\pi$ for $s\to\infty$.

For the $S$-wave, we will take into account the two-channel
rescattering effects. Along the right-hand cut, the two-channel
unitarity conditions reads
\begin{equation}\label{eq.unitarity2channel}
\textrm{Im}\, \vec{M}_0(s)=2i T_0^{0\ast}(s)\Sigma(s)
\left[\vec{M}_0(s)+\hat{\vec{M}}_0(s)\right] \,,
\end{equation}
where the two-dimensional vectors $\vec{M}_0(s)$ and
$\hat{\vec{M}}_0(s)$ contain the right-hand-  and the
left-hand-cut parts of both the $\pi\pi$ and the $K\bar{K}$
final states, respectively,
 \begin{equation}
\vec{M}_0(s)=\left( {\begin{array}{*{2}c}
   {M^\pi_0(s)} \\
   {\frac{2}{\sqrt{3}}M^K_0(s)}  \\\end{array}} \right), \hspace{0.5cm}\hat{\vec{M}}_0(s)=\left( {\begin{array}{*{2}c}
   {\hat{M}^\pi_0(s)} \\
   {\frac{2}{\sqrt{3}}\hat{M}^K_0(s)}  \\
\end{array}} \right)\,.
 \end{equation}
The two-dimensional matrices $T_0^0(s)$ and $\Sigma(s)$ are
\begin{equation}\label{eq.T00}
T_0^0(s)=
 \left( {\begin{array}{*{2}c}
   \frac{\eta_0^0(s)e^{2i\delta_0^0(s)}-1}{2i\sigma_\pi(s)} & |g_0^0(s)|e^{i\psi_0^0(s)}   \\
  |g_0^0(s)|e^{i\psi_0^0(s)} & \frac{\eta_0^0(s)e^{2i\left(\psi_0^0(s)-\delta_0^0(s)\right)}-1}{2i\sigma_K(s)} \\
\end{array}} \right)
\end{equation}
and
$\Sigma(s)\equiv \text{diag} \big(\sigma_\pi(s)\theta(s-4m_\pi^2),\sigma_K(s)\theta(s-4m_K^2)\big)$.
There are three input functions in the $T_0^0(s)$ matrix: the
$\pi\pi$ $S$-wave isoscalar phase shift $\delta_0^0(s)$, for which
we will use the result from the Roy equation analysis in
Ref.~\cite{Leutwyler2012}; the $\pi\pi \to K\bar{K}$ $S$-wave
amplitude $g_0^0(s)=|g_0^0(s)|e^{i\psi_0^0(s)}$ with modulus and
phase, for which the results based on the Roy--Steiner approach in
Ref.~\cite{Moussallam2004} will be used. These inputs are used below
the appearance of additional inelasticities from $4\pi$ intermediate states,
namely up to $\sqrt{s_0}=1.3\GeV$ (the $f_0(1370)$ resonance is known to have a significant coupling to $4\pi$~\cite{Olive:2016xmw}).
Above $s_0$, the phases
$\delta_0^0(s)$ and $\psi_0^0$ are guided smoothly to
2$\pi$~\cite{Moussallam2000}
\begin{equation}
\delta(s)=2\pi+(\delta(s_0)-2\pi)\frac{2}{1+(\frac{s}{s_0})^{3/2}}\,.
\end{equation}
The inelasticity $\eta_0^0(s)$ in Eq.~\eqref{eq.T00} is related to
the modulus $|g_0^0(s)|$ by
\begin{equation}
\eta_0^0(s)=\sqrt{1-4\sigma_\pi(s)\sigma_K(s)|g_0^0(s)|^2\theta(s-4m_K^2)}\,.
\end{equation}
The numerical solution of the homogeneous two-channel unitarity
relation
\begin{equation}\label{eq.unitarity2channelhomo}
\textrm{Im}\, \Omega(s)=2i T_0^{0\ast}(s)\Sigma(s) \Omega(s),
\hspace{1cm}  \Omega(0)=\mathbbm{1} \,,
\end{equation}
has been computed in
Refs.~\cite{Leutwyler90,Moussallam2000,Hoferichter:2012wf,Daub}.
Using $\Omega(s)$, the solution of the inhomogeneous two-channel
unitarity condition in Eq.~\eqref{eq.unitarity2channel} is given by
\begin{equation}\label{OmnesSolution2channel}
\vec{M}_0(s)=\Omega(s)\bigg\{\vec{P}^{n-1}(s)+\frac{s^n}{\pi}\int_{4m_\pi^2}^\infty
\frac{\diff x}{x^n}\frac{\Omega^{-1}(x)T(x)\Sigma(x)\hat{\vec{M}}_0(x)}{x-s}\bigg\} \,.
\end{equation}

To determine the required number of subtractions that guarantees the
dispersive integrals in Eqs.~\eqref{OmnesSolution1channel}
and~\eqref{OmnesSolution2channel} to converge, we need to
investigate the high-energy behavior of the integrand. First it is
known that for the single-channel Omn\`es function defined in
Eq.~\eqref{Omnesrepresentation}, it falls off asymptotically as
$s^{-k}$ if the phase shift $\delta_l^I(s)$ approaches $k\pi$ at
high energies. Since the $D$-wave $\pi\pi$ phase shift,
$\delta_2^0(s)$, reaches $\pi$ for high energies, we have
$\Omega_2^0(s)\sim 1/s$ for large $s$. Second, the high-energy
behavior of the two-channel Omn\`es function has been analyzed in
Ref.~\cite{Moussallam2000}, and the $1/s$ asymptotic behavior of
$\Omega_l^I(s)$ is ensured by the asymptotic condition $\sum
\delta_l^I(s) \geq 2\pi$ for $s\to\infty$, where $\sum
\delta_l^I(s)$ is the sum of the eigen phase shifts. Third, we have
checked that in an intermediate energy range of $1\GeV^2 \lesssim s
\ll m_{\Upsilon}^2$, both the inhomogeneities contributed by the
$Z_b$-exchange term and the box graphs term grow at most linearly in
$s$. So we conclude that in the dispersive representations for
$M_2(s)$ and $\vec{M}_0(s)$, three subtractions are sufficient to
make the dispersive integrals convergent.

At low energies, the amplitudes $M_2(s)$ and $\vec{M}_0(s)$ should
match to the results of chiral perturbation theory. Namely, in the
limit of switching off the final-state interactions,
$\Omega_2^0(s)=1$ and $\Omega(0)=\mathbbm{1}$, the subtraction
functions agree with the chiral representations given in
Eqs.~\eqref{eq.M0chiral} and~\eqref{eq.M2chiral}.  Since both
$M_0^\chi(s)$ and $M_2^\chi(s)$ grow no faster than $\sim s^2$, they
can be covered by the degree of the subtractions. Therefore, for the
$D$-wave, the integral equation takes the form \be\label{M21channel}
M_2^\pi(s)=\Omega_2^0(s)\bigg\{M_2^{\chi,\pi}(s)+\frac{s^3}{\pi}\int_{4m_\pi^2}^\infty
\frac{\diff x}{x^3} \frac{\hat
M_2^\pi(x)\sin\delta_2^0(x)}{|\Omega_2^0(x)|(x-s)}\bigg\} \,. \ee
For the $S$-wave, the integral equation reads
\begin{equation}\label{M02channel}
\vec{M}_0(s)=\Omega(s)\bigg\{\vec{M}_0^{\chi}(s)+\frac{s^3}{\pi}\int_{4m_\pi^2}^\infty
\frac{\diff x}{x^3}\frac{\Omega^{-1}(x)T(x)\Sigma(x)\hat{\vec{M}}_0(x)}{x-s}\bigg\} \,,
\end{equation}
where $ \vec{M}^{\chi}_0(s)=\big(
   M_0^{\chi,\pi}(s),
   2/\sqrt{3}\,M_0^{\chi,K}(s)
   \big)^{T}$.

The differential decay width
for $\Upsilon(mS) \rightarrow \Upsilon(nS) \pi^+\pi^-$
with respect to the $\pi\pi$ invariant
mass and the helicity angle reads
\begin{equation}
\frac{\diff\Gamma}{\diff \sqrt{s} \,\diff\cos\theta} =
\frac{\sqrt{s}\,\sigma_\pi |\vec{q}|}{128\pi^3 m_{\Upsilon(mS)}^2}
\left|M_0^\pi+\hat{M}_0^\pi+(M_2^\pi+\hat{M}_2^\pi)
P_2(\cos\theta)\right|^2\,.\label{eq.pipimassdistribution}
\end{equation}

\section{Phenomenological discussion}\label{pheno}

The experimental data considered in this work are the $\pi\pi$
invariant mass distributions of the $\Upsilon(4S) \rightarrow
\Upsilon(1S,2S) \pi^+\pi^-$ decays measured by the BaBar~\cite{BABAR2006} and
Belle Collaborations~\cite{Belle2009}.

The chiral coupling constants $c_i$ in
Eq.~\eqref{LagrangianUpUppipi} are different for the two decays $\Upsilon(4S)
\rightarrow \Upsilon(1S) \pi^+\pi^-$ and $\Upsilon(4S)
\rightarrow \Upsilon(2S) \pi^+\pi^-$, since there is no
symmetry connecting the bottomonium states with different radial
excitations. The mass difference between the two $Z_b$ states is
much smaller than the difference between their masses and the
$\Upsilon(lS)\pi$ $(l=1,2)$ thresholds as well as
$m_{\Upsilon(4S)}-m_\pi$; they have the same quantum numbers and
thus the same coupling structure as given by
Eq.~\eqref{LagrangianZbUppi}. So the $Z_{b}(10610)$ and $Z_b(10650)$ contributions
are strongly correlated in the fit, and it is very difficult to
distinguish their effects from each other in the processes studied
in this work. Therefore we only use one $Z_b$, the
$Z_{b}(10610)$, in our fit by setting $C_{nm,2}=0$ as in our
previous analysis of
$\Upsilon(3S)\to\Upsilon(1,2S)\pi\pi$~\cite{Chen2016}. For the input
mass of the $Z_{b}(10610)$, we will use the heavy-quark spin symmetry
conserving result given in Ref.~\cite{Guo2016}. The value of
$g_{JHH(4S)}$ is extracted from the measured open-bottom decay widths of the
$\Upsilon(4S)$, $g_{JHH(4S)}=1.43\GeV^{-3/2}$.

For each $\Upsilon(4S) \to \Upsilon(nS) \pi\pi$ $(n=1, 2)$
process, the unknown parameters are $c_1$ and $c_2$ corresponding
to the chiral contact $\Upsilon\Upsilon^{\prime}\Phi\Phi$ coupling,
$C_{4n,1}$ related to the $Z_b$ exchange mechanism, and
$g_{JHH(nS)}$ for the box diagrams. To illustrate the
effects of the $Z_b$-exchange and the box graph mechanisms, we
perform several fits by choosing different strategies. Fit~I only
includes the chiral contact $c_i$ terms; Fit~II adds the
$Z_b$-exchange terms to them. Fit~III includes the chiral contact
$c_i$ terms and the box diagrams, and finally Fit~IV takes all of
the contact $c_i$ terms, the $Z_b$ exchange, and the box diagrams
into account. FSI is included in all fits.
\begin{figure}
\centering
\includegraphics*[width=0.499\linewidth]{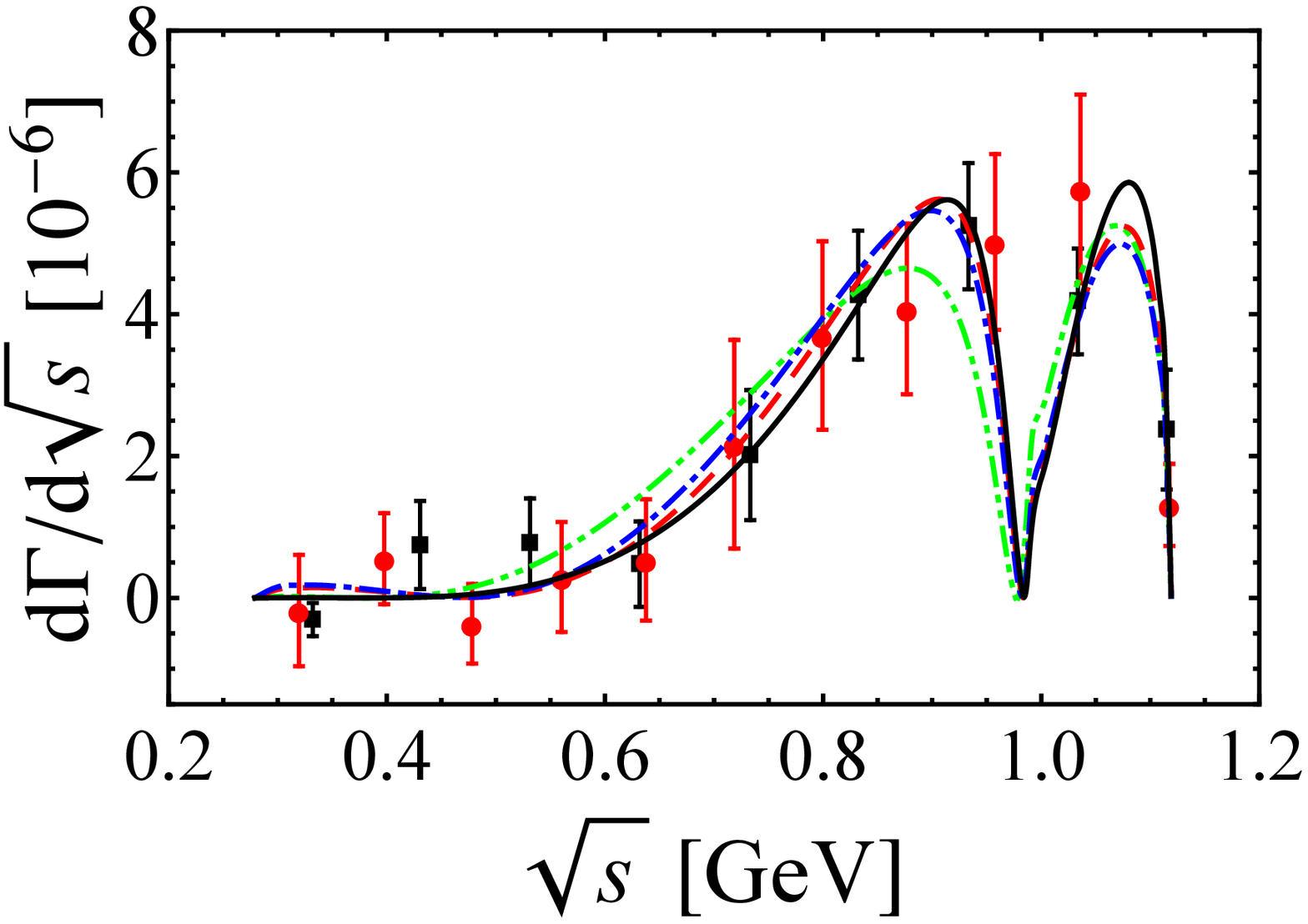}\hfill
\includegraphics*[width=0.499\linewidth]{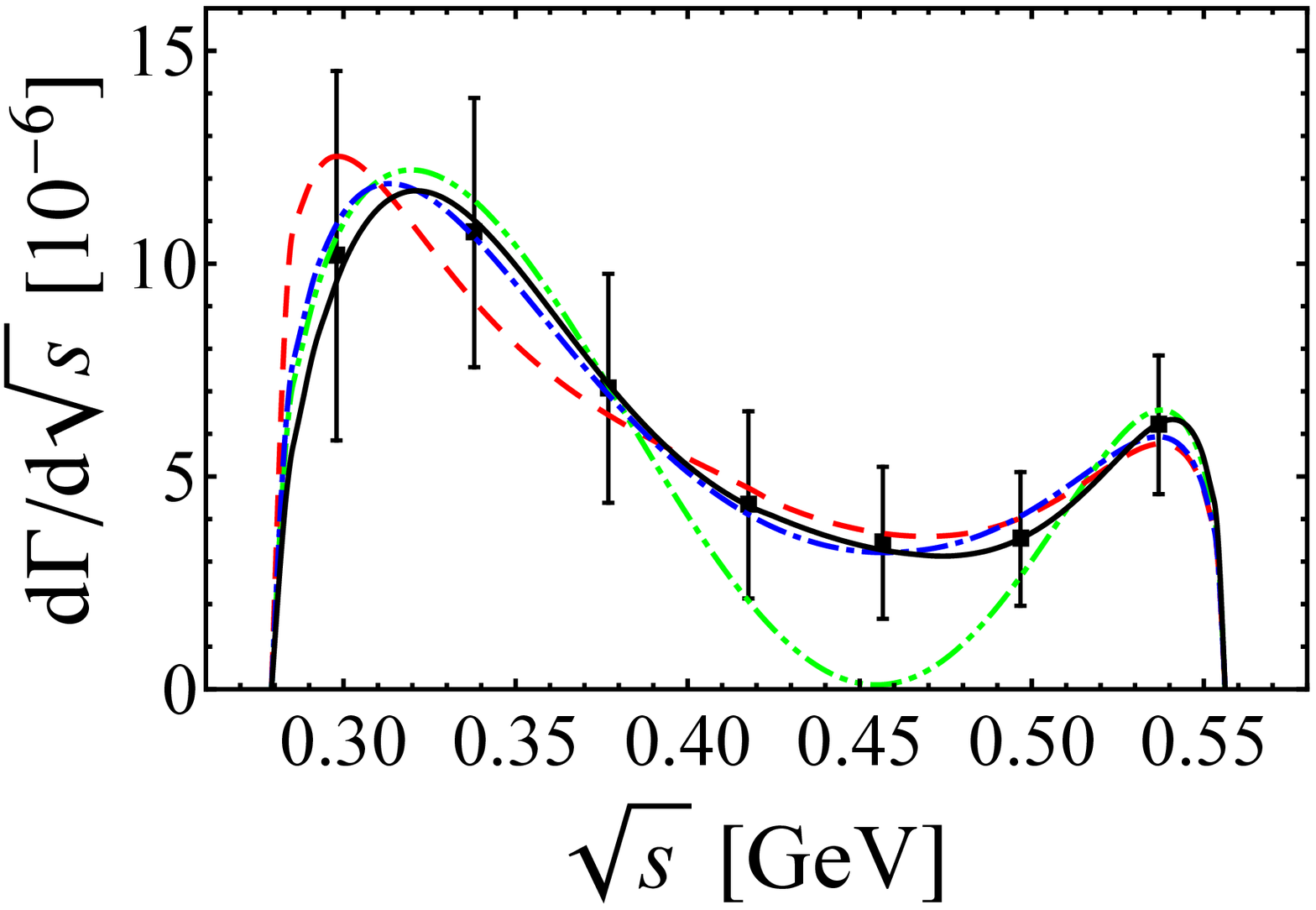}
\caption{Fit results of the $\pi\pi$ invariant mass spectra for the
decays $\Upsilon(4S) \to \Upsilon(1S) \pi^+\pi^- $ (left) and $\Upsilon(4S)
\to \Upsilon(2S) \pi^+\pi^- $ (right). The solid squares and solid circles
denote the data from the BaBar Collaboration~\cite{BABAR2006} and
Belle Collaboration~\cite{Belle2009}, respectively. Fit~I (green
dash-dot-dotted): only including the chiral contact terms $c_i$,
Fit~II (red dashed):  chiral contact terms and the $Z_b$-exchange
term, Fit~III (blue dot-dashed): chiral contact terms and box
diagrams, Fit~IV (black solid): including the contact terms $c_i$,
the $Z_b$-exchange term, and the box diagrams. FSI is included in
all fits.}\label{fig.pipimassdistribution}
\end{figure}
In Fig.~\ref{fig.pipimassdistribution}, the
fitted results of Fits~I--IV are shown as
the green dash-dot-dotted, red dashed, blue dot-dashed, and black
solid lines, respectively.
\begin{table}
\caption{\label{tablepar1} Fit parameters from the best fits
of the $\Upsilon(4S) \to \Upsilon(nS) \pi\pi$ $(n=1, 2)$ processes.}
\renewcommand{\arraystretch}{1.2}
\begin{center}
\begin{tabular}{l|cc}
\toprule
         & $~\Upsilon(4S) \to \Upsilon(1S) \pi^+\pi^- ~$
         & $~\Upsilon(4S) \to \Upsilon(2S) \pi^+\pi^- ~$\\
\hline
$c_1~[\text{GeV}^{-1}]$   &   $ (9.8\pm 1.0)\times 10^{-4}$           & $(1.2\pm 0.6)\times 10^{-1}$ \\
$c_2~[\text{GeV}^{-1}]$   &   $ (-1.6\pm 1.1)\times 10^{-4}$  & $(-1.0\pm 0.6)\times 10^{-1}$  \\
$ C_{4n,1}$ & $ (2.6\pm 1.3)\times 10^{-4}$ &$ (-3.2\pm 1.8)\times 10^{-2}$ \\
$g_{JHH(nS)}~[\text{GeV}^{-\frac{3}{2}}]$ & $(8.6\pm 6.1)\times 10^{-5}$ &$(1.7\pm 0.8)\times 10^{-2}$ \\
\hline
 ${\chi^2}/{\rm d.o.f}$ &  ${10.45}/{(20-4)}=0.65$   &  ${0.04}/{(7-4)}=0.01$  \\
\botrule
\end{tabular}
\end{center}
\renewcommand{\arraystretch}{1.0}
\end{table}
The fitted parameters as well as the $\chi^2/\text{d.o.f.}$ of our
best fit, Fit~IV, are shown in Table~\ref{tablepar1}. We find very
different values for the parameters $c_1$ and $c_2$ from fitting the
data of transitions between different $\Upsilon(lS)$ states. These
low-energy constants parameterize the nonperturbative QCD matrix
elements of gluonic operators between the initial and final
bottomonia. For different initial and final $\Upsilon$ states, these
parameters are not related to each other at the hadronic level, and
can well be very different. In principle, the parameter values from
the fit in this paper cannot be directly compared with those in
Ref.~\cite{Chen2016}, which do not include the box diagrams when
analyzing the $\Upsilon(3S)$ and $\Upsilon(2S)$ dipion transitions.
We thus made a new fit to the decay
$\Upsilon(3S)\to\Upsilon(1S)\pi\pi$ studied therein. It turns out
that the values of $c_1$ and $c_2$ decrease only by around 35\% in
comparison with those given in Table~I of Ref.~\cite{Chen2016}. Our
fittings turn out to indicate the following hierarchy:
$|c_{1,2}^{4\to1}| \ll |c_{1,2}^{4\to2}| \lesssim|c_{1,2}^{3\to1}|
\ll |c_{1,2}^{3\to2}|$. This may be understood from the node
structure of the $\Upsilon$ wave functions: for the processes with
the same initial $\Upsilon$ state, the larger the difference between
the principal quantum numbers, the smaller the gluonic matrix
elements and thus the magnitude of the parameters. Note that the
total $\chi^2$ value for the transition $\Upsilon(4S) \to
\Upsilon(2S) \pi^+\pi^-$ is very low, $\chi^2/\text{d.o.f.} = 0.01$.
This small number reflects the observation that the fluctuation in
the data appears to be significantly smaller than what the error
bars allow for, which indicates that they might well be dominated by
systematics.

\begin{figure}
\centering
\includegraphics[width=\linewidth]{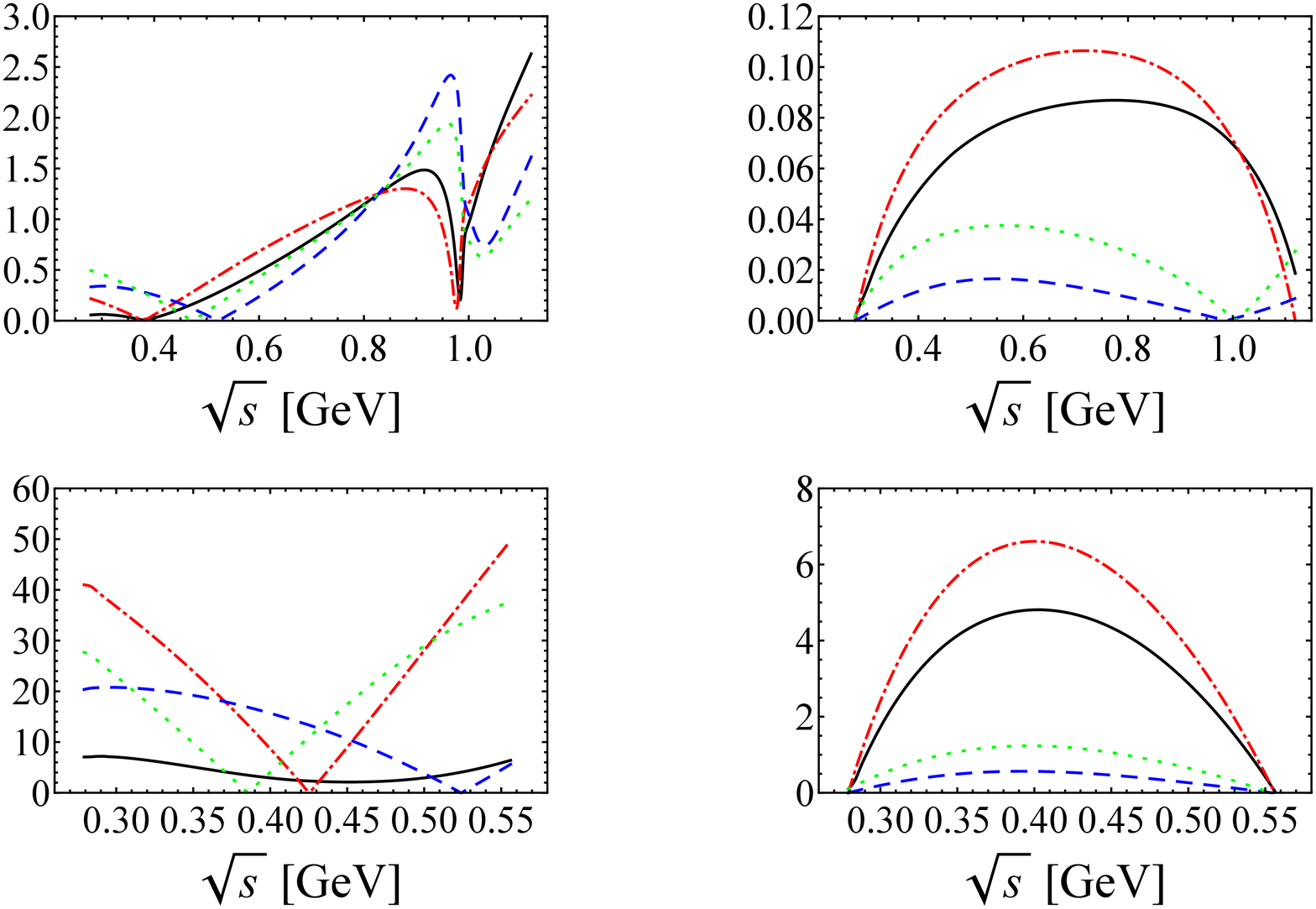}
\caption{Moduli of the $S$- (left) and $D$-wave (right) amplitudes
in $\Upsilon(4S) \to \Upsilon(1S) \pi^+\pi^-$ (top) and
$\Upsilon(4S) \to \Upsilon(2S) \pi^+\pi^-$ (bottom). The black solid
lines represent our best fit results, while the red dot-dashed, blue
dashed, and green dotted lines correspond to the contributions from
the $c_i$ terms, the $Z_b(10610)$, and the box diagrams,
respectively. }\label{fig.SwaveDwaveAmplitudes}
\end{figure}

\begin{figure}[tb]
\centering
\includegraphics*[width=1\linewidth]{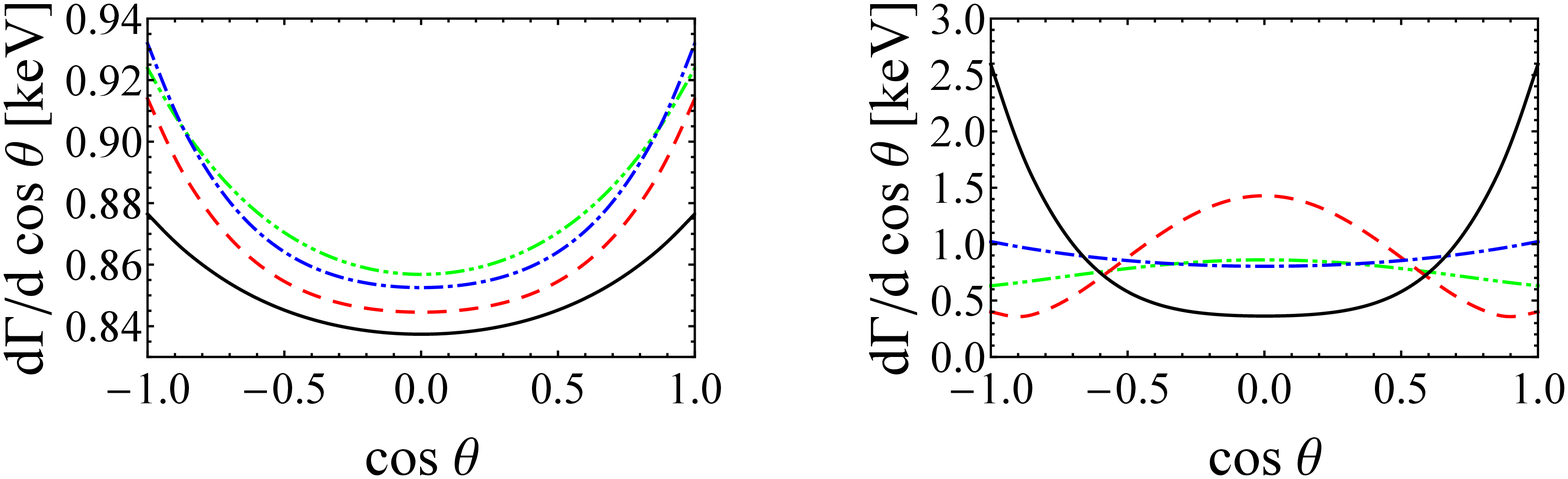}
\caption{Theoretical predictions of the helicity angular
distributions for the decays $\Upsilon(4S) \to \Upsilon(1S)
\pi^+\pi^- $ (left) and $\Upsilon(4S) \to \Upsilon(2S) \pi^+\pi^- $
(right). The line style is as in
Fig.~\ref{fig.pipimassdistribution}.
}\label{fig.angulardistribution}
\end{figure}

Using the central values of the parameters in the best fit, in
Fig.~\ref{fig.SwaveDwaveAmplitudes} we plot the moduli of the $S$-
and $D$-wave amplitudes from the $c_i$ terms, the $Z_b(10610)$
state, and the box graphs for the processes $\Upsilon(4S) \to
\Upsilon(1S) \pi^+\pi^-$ and $\Upsilon(4S) \to \Upsilon(2S)
\pi^+\pi^-$, respectively. In addition, in
Fig.~\ref{fig.angulardistribution} we show the resulting theoretical
predictions for the angular distributions.

As shown in Fig.~\ref{fig.pipimassdistribution}, including the
$Z_b$-exchange and the box graph contributions improves the fit
quality for $\Upsilon(4S) \to \Upsilon(1S) \pi^+\pi^-$ only
marginally, mainly in the region around $1\GeV$. However, for
$\Upsilon(4S) \to \Upsilon(2S) \pi^+\pi^-$, the fit quality
increases significantly when considering either of those two
mechanisms (or both). Loop effects were already studied in the
${}^3P_0$ quark-pair-creation model in Ref.~\cite{Kuang1991}, and
found to be tiny for $\Upsilon(3S,2S) \to \Upsilon(2S,1S)
\pi^+\pi^-$. This is probably due to the fact that $\Upsilon(3S,2S)$
are too far below the $B\bar{B}$ threshold. This situation is
expected to change for the $\Upsilon(4S)$, with the open-bottom
channels contributing significantly to its decay rate. In
Fig.~\ref{fig.SwaveDwaveAmplitudes}, one observes that for the
dominant $S$-wave amplitudes, the contributions from the $c_i$
terms, from the $Z_b$-exchange term, and from the box diagram term
are all of the same order. Especially, for the decay $\Upsilon(4S)
\to \Upsilon(1S) \pi^+\pi^-$, the box graphs and the $Z_b$ exchange
play a major role in the energy range around $0.95\GeV$, and account
for the better description of the data there. Note that the
contribution of loops including $B_s$ mesons, producing kaons that
subsequently rescatter into a pion pair, is entirely negligible: in
the NREFT formalism, these graphs vanish at the $K\bar K$ threshold.
For the $D$-wave, the contributions from $Z_b$ exchange and the box
graphs are much smaller than that from the $c_i$ terms. We should
mention that the plots in Fig.~\ref{fig.SwaveDwaveAmplitudes}
correspond to using the central values of the best fit parameters.
The shapes of the curves corresponding to the box diagrams and the
$Z_b$-exchange terms are similar; however, their relative strength
is not very meaningful because there is a strong correlation in the fit
between the parameters $C_{41,1}$ and $J_{JHH(1S)}$. This can be
easily seen from the fact that the curves for Fit~II and Fit~III are
very similar to each other in
Fig.~\ref{fig.pipimassdistribution}~(left), which means that the
$Z_b$-exchange and box terms can hardly be distinguished in the
$\pi\pi$ invariant mass distribution of the transition $\Upsilon(4S)
\to \Upsilon(1S) \pi^+\pi^-$.

Notice that in Refs.~\cite{Meng2008,DYChen2011}, the loop
contribution of the sequential process $\Upsilon(4S) \to B\bar{B}
\to \Upsilon(nS)S \to \Upsilon(nS) \pi^+\pi^- (n=1,2)$, where the
scalar $S$ can correspond to the $f_0(500)$ and the $f_0(980)$, has
been considered. This kind of loop topology can be described by
Fig.~\ref{fig.FeynmanDiagramTriangle}\,(a) including FSIs, which is
suppressed compared to the box graphs in NREFT. In our scheme, the
FSIs are taken into account in a model-independent way, and we do not
have to specify the contributing scalar resonances. Another merit of
our calculation is that, instead of only obtaining the absorptive
part of the loops by using Cutkosky
rules~\cite{Meng2008,DYChen2011}, we completely compute both their real
and imaginary parts.

An interesting feature of the $\pi\pi$ invariant mass distribution of
$\Upsilon(4S) \to \Upsilon(1S) \pi^+\pi^-$ is that the older Belle data
from Ref.~\cite{Belle2005} hint at a two-peak structure in the range of
$m_{\pi^+\pi^-}=0.8 \ldots 1.2\GeV$, while the later measurements given in
Refs.~\cite{BABAR2006,Belle2009} do not display such a feature in any obvious
way. As the mass difference between $\Upsilon(4S)$ and $\Upsilon(1S)$ is about
$1.12\GeV$, the isoscalar-scalar $f_0(980)$ meson, which couples strongly to
$\pi\pi$, should be visible in the spectrum. With FSI described reliably in the
dispersive approach, we see that the $f_0(980)$ indeed accounts for a dip at its
mass, and a two-peak structure is naturally produced.
A possible reason why such a two-peak structure is not observed in
Refs.~\cite{BABAR2006,Belle2009} may be the wide energy bins used in these
experimental measurements. The fact that the $f_0(980)$ should be manifest in the
$\pi\pi$ invariant mass distribution of $\Upsilon(4S) \to \Upsilon(1S)
\pi^+\pi^-$ has already been emphasized in Ref.~\cite{Guo2007:4S}.
The dip caused by the $f_0(980)$ is also present in the
calculation of Ref.~\cite{Surovtsev:2015hna}.

For the $\Upsilon(4S) \to \Upsilon(2S) \pi^+\pi^-$ process, it is
known that the two-hump behavior in the $\pi\pi$ invariant mass
spectra is incompatible with the prediction from the QCD multipole
expansion, resembling the case of $\Upsilon(3S) \to \Upsilon(1S)
\pi\pi$~\cite{Kuang1991,Eichten2008,Chen2016}. In the formalism
outlined above, the original formulation of the QCD multipole expansion appears by
including only the tree-level $c_i$-terms, however, omitting the $\pi\pi$
FSIs.
As shown by the blue dot-dashed line in the right panel of
Fig.~\ref{fig.pipimassdistribution}, including
the final-state interaction can roughly reproduce a two-hump structure. However, it
produces a zero in the amplitude inside the physical region and the
agreement with the data is not very convincing. This feature was
also observed in our previous study of $\Upsilon(3S) \to
\Upsilon(1S) \pi\pi$, where, however, a simultaneous fit of the
$\pi\pi$ invariant mass and the helicity angular distributions
cannot reproduce the two-hump behavior in the dipion mass spectra by
only using the $c_i$ terms~\cite{Chen2016}. The angular distribution
data are therefore important to distinguish the effects of different
mechanisms. In Fig.~\ref{fig.angulardistribution}, the theoretical
predictions of the helicity angular distributions in different fit
scenarios are shown. For $\Upsilon(4S) \to \Upsilon(2S) \pi^+\pi^-$,
the angular distributions are distinctly different when including
the $Z_b$-exchange and box graphs terms, hence these results can be
used to check their effects when experimental data become available
in the future.

\begin{figure}
\centering
\includegraphics[width=\linewidth]{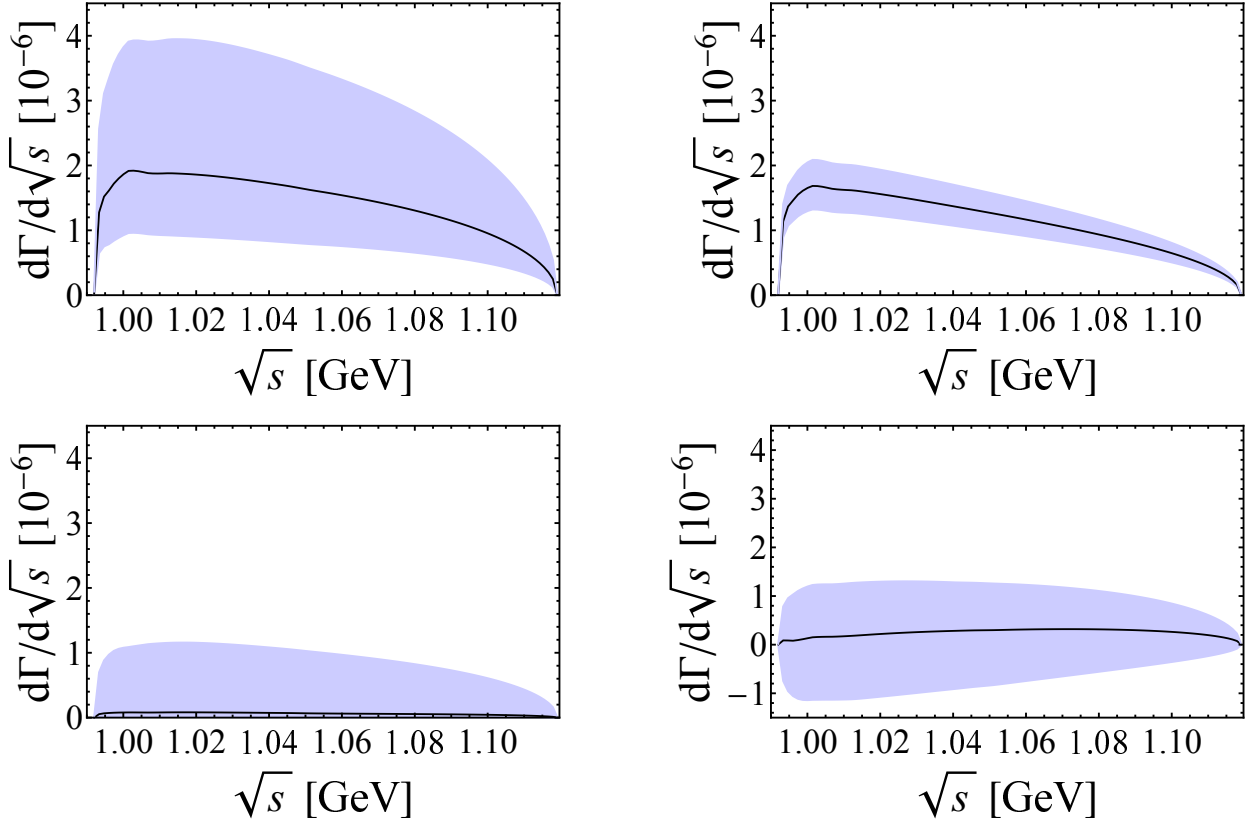}
\caption{Theoretical prediction of the $K\bar{K}$ invariant mass
spectrum for the decay $\Upsilon(4S) \to \Upsilon(1S) K^+K^-$ (top
left). The contributions from the $c_i$ terms (top right), the
$Z_b(10610)$ state plus box graphs (bottom left), and their
interference (bottom right) are also depicted. The shaded areas
corresponds to the error band.}\label{fig.KKmassdistribution}
\end{figure}

Using the fit parameters given in Table~\ref{tablepar1}, we can
predict the decay width of $\Upsilon(4S) \rightarrow \Upsilon(1S)
K^+ K^-$, as well as the corresponding $K\bar K$ invariant mass
distribution. The relevant Feynman diagrams can be obtained by
replacing all external pions by kaons in
Fig.~\ref{fig.FeynmanDiagram}, but without diagram~(b1) due to the
absence of a $Z_b\Upsilon K$ vertex. The $Z_b$ contributes also to
$K\bar K$ through diagram~(b2) due to the final-state interactions
that, especially around the $K\bar K$ threshold, provides strong
$\pi\pi\to K\bar K$ transitions. Most ingredients of the amplitude
of the $\Upsilon(4S) \rightarrow \Upsilon(1S) K^+ K^-$ process have
been given in Sec.~\ref{theor}.  We omit the $K\bar K$ $D$-wave,
which is negligible due to its strong near-threshold suppression.
Within $1\sigma$ uncertainties, the prediction of the decay width of
$\Upsilon(4S) \rightarrow \Upsilon(1S) K^+ K^-$ is
\begin{equation}
\Gamma_{\Upsilon(4S) \rightarrow \Upsilon(1S) K^+K^-}=0.18_{-0.09}^{+0.21} \,\text{keV}\,,\label{eq.DecayWidthOf4S1SKK}
\end{equation}
corresponding to a branching fraction of
$0.9^{+1.0}_{-0.4}\times10^{-5}$, and the dikaon invariant mass
spectrum is given in Fig.~\ref{fig.KKmassdistribution} (top left).
The rapid rise of the $K\bar K$ invariant mass distribution in the
near-threshold region is a result of the $f_0(980)$, in line with
the dip around $1\GeV$ in Fig.~\ref{fig.pipimassdistribution}. Like
the $\Upsilon(4S) \rightarrow \Upsilon(1S) \pi^+ \pi^-$ process,
there is a strong correlation between the $Z_b$-exchange terms and
the box diagrams in the $\Upsilon(4S) \rightarrow \Upsilon(1S) K^+
K^-$ process, and in Fig.~\ref{fig.KKmassdistribution} we also plot
the contributions from the the $c_i$ terms (top right), the
$Z_b(10610)$ state plus box graphs (bottom left), and their
interference (bottom right), respectively. One finds that for the
central values of the theoretical predictions, the $Z_b$-exchange
term and the box graphs nearly cancel each other, and the total line
shape is quite similar to the $c_i$ terms only. Both the rapid rise
in the $m_{K\bar K}$ distribution and the nontrivial structure in
the large $m_{\pi\pi}$ region of the dipion invariant mass
distribution are due to the final-state interactions between the
light mesons, depicted in
Fig.~\ref{fig.FeynmanDiagram}~(c1,\,d1,\,a2,\,\ldots,\,d2), which
receive contributions from both the $Z_b$-exchange and box diagrams.
As a result, their strong correlation in the fit to the data of the
dipion transitions leads to the significant cancellation in the
prediction of the $m_{K\bar K}$ distribution. The large spread
mainly comes from the uncertainties of $Z_b(10610)$ plus box graphs,
and the interference term. These predictions encourage future
experimental measurements in this channel.

\section{Conclusions}
\label{conclu}

We have studied the effects of $Z_b$ exchange and bottom meson loops
in the decays $\Upsilon(4S) \to \Upsilon(nS) \pi\pi$ $(n=1, 2)$. The
bottom meson loops are treated in the NREFT scheme, in which the
power counting rules indicate that the box diagrams are dominant.
The strong FSIs, especially the coupled-channel FSI in the $S$-wave,
are taken into account model-independently by using dispersion
theory. The forms of the subtraction functions are obtained by
matching to the leading chiral contact terms. Through fitting the
data of the $\pi\pi$ invariant mass spectra, the couplings of the
$\Upsilon\Upsilon^{\prime}\pi\pi$ and $\Upsilon B^{(*)}B^{(*)}$
vertices, as well as the product of couplings of the
$Z_b\Upsilon\pi$ and $Z_b \Upsilon^\prime\pi$ vertices are
determined (where $\Upsilon$ and $\Upsilon'$ denote the final- and
initial-state bottomonia). For the dominant $S$-wave component, it
is found that the contributions from $Z_b$ exchange, the loops, and
the chiral contact term are of the same order. For $\Upsilon(4S) \to
\Upsilon(2S) \pi^+\pi^-$, including the $Z_b$-exchange term and the
bottom meson loops naturally describes the two-hump behavior in the
$\pi\pi$ invariant mass distribution. Unfortunately, the present
data are insufficient to distinguish between the effects of the
$Z_b$ exchange and the bottom meson loops.  We provide theoretical
predictions of the helicity angular distributions, which may be
useful to identify the effects of $Z_b$-exchange and bottom meson
loops with future experimental data. For the $\Upsilon(4S) \to
\Upsilon(1S) \pi^+\pi^-$ decay, we expect that there is a dip in the
$\pi\pi$ spectrum around $1\GeV$, caused by the opening of the
$K\bar{K}$ channel near the $f_0(980)$ resonance. This dip has
probably not been observed yet in the present experimental data yet
due to lack of sufficiently precise energy resolution. Improved data
to resolve this issue is eagerly awaited. We also predict the decay
width and the $K\bar{K}$ invariant mass distribution of the
$\Upsilon(4S) \rightarrow \Upsilon(1S) K^+ K^-$ process,
demonstrating the usefulness of this additional measurement that
should be feasible at Belle~II.

\section*{Acknowledgments}

We are grateful to Zhi-Hui Guo, Claudia Patrignani, and Qian Wang
for helpful discussions, and to Roman Mizuk for the proposal to add
the investigation of the $K^+K^-$ final state. This work is
supported in part by NSFC and DFG through funds provided to the
Sino--German Collaborative Research Center (CRC)110 ``Symmetries and
the Emergence of Structure in QCD'' (NSFC Grant No.~11621131001, DFG
Grant No.~TRR110), by NSFC (Grant No.~11647601), by the Thousand
Talents Plan for Young Professionals, by the CAS Key Research
Program of Frontier Sciences (Grant No.~QYZDB-SSW-SYS013), and by
the CAS President's International Fellowship Initiative (PIFI)
(Grant No.~2015VMA076). MC also acknowledges support by the Spanish
Ministerio de Economia y Competitividad (MINECO) under the project
MDM-2014-0369 of ICCUB (Unidad de Excelencia 'Mar\'\i a de Maeztu'),
and, with additional European Fondo Europeo de Desarrollo Regional
(FEDER) funds, under the contract FIS2014-54762-P as well as support
from the Ge\-ne\-ra\-li\-tat de Catalunya contract 2014SGR-401, and
from the Spanish Excellence Network on Hadronic Physics
FIS2014-57026-REDT.
%
\begin{appendix}
%
\section{Remarks on the box diagrams and four-point integrals}\label{app:box}
%

In this appendix, we will discuss the calculation of the amplitudes that involve
four-point loop integrals in some detail. We will start by discussing the
parametrization and simplification of scalar four-point integrals. Then we will
introduce a tensor reduction scheme to deal with higher-rank integrals. Finally,
we will give the leading part of the corresponding integrals (proportional to
$\bm{\epsilon}_{\Upsilon'}\cdot\bm{\epsilon}_\Upsilon$) for the possible
intermediate bottom mesons.

\subsection{Scalar four-point integrals}
\begin{figure}[t]
\begin{center}
\includegraphics[width=0.9\linewidth]{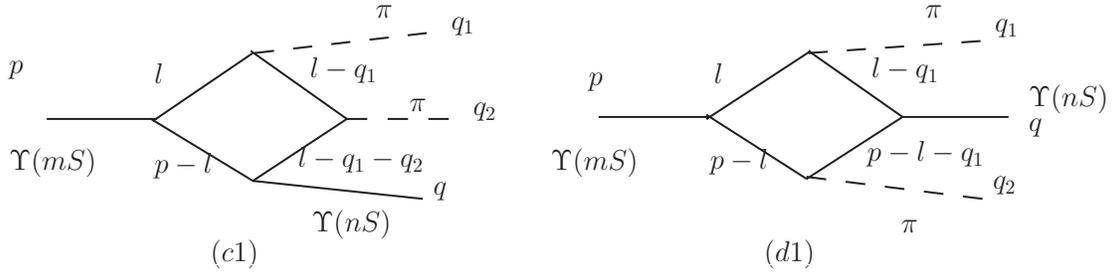}
\caption{Kinematics used in the calculation of the four-point integrals. }
\label{fig:BoxLabel}
\end{center}
\end{figure}
Because of the simpler structure we begin with the first topology as shown in Fig.~\ref{fig:BoxLabel}. The corresponding scalar integral, evaluated for the initial bottomonium at rest ($p=(M,\vec 0)$) and labelled $J^{(0c)}$ to be consistent with Fig.~\ref{fig.FeynmanDiagram}, reads
{\allowdisplaybreaks
\begin{align}
 J^{(0c)}&\equiv i\int \!\frac{d^4l}{(2\pi)^4}\frac{1}{[l^2-m_1^2+i\epsilon][(p-l)^2-m_2^2+i\epsilon][(l-q_1-q_2)^2-m_3^2+i\epsilon][(l-q_1)^2-m_4^2+i\epsilon]}\nonumber\\
&\simeq\frac{-i}{16m_1m_2m_3m_4}\int \! \frac{d^4l}{(2\pi)^4}
\frac{1}{\left[l^0-\frac{\vec l^2}{2m_1}-m_1+i\epsilon\right]\left[l^0-M+\frac{\vec l^2}{2m_2}+m_2-i\epsilon\right]} \nonumber\\
&\times \frac{1}{\left[l^0-q_1^0-q_2^0-\frac{(\vec l+\vec q)^2}{2m_3}-m_3+i\epsilon\right]\left[l^0-q_1^0-\frac{(\vec l-\vec q_1)^2}{2m_4}-m_4+i\epsilon\right]} .
\end{align}}%
Performing the contour integration is straightforward since only one
pole is located in the upper half-plane. We find
\begin{equation}
-\frac{\mu_{12}\mu_{23}\mu_{24}}{2m_1m_2m_3m_4}\int \! \frac{d^3l}{(2\pi)^3}\frac{1}{
[\vec l^2+c_{12}-i\epsilon][\vec l^2+2\frac{\mu_{23}}{m_3}\vec l\cdot \vec q+c_{23}-i\epsilon][\vec l^2-2\frac{\mu_{24}}{m_4}\vec l\cdot \q_1+c_{24}-i\epsilon]} , \label{eq:topo1}
\end{equation}
where we defined
\begin{align}
 c_{12} &\equiv 2\mu_{12}\left(m_1+m_2-M\right), \quad c_{23}\equiv 2\mu_{23}\left(m_2+m_3-M+q_1^0+q_2^0+\frac{\q^2}{2m_3}\right), \nonumber\\
 c_{24} &\equiv 2\mu_{24}\left(m_2+m_4-M+q^0_1+\frac{\q_1^2}{2m_4}\right), \quad \mu_{ij}=\frac{m_im_j}{m_i+m_j}.
\end{align}

For the second topology we immediately find
\begin{align}
 J^{(0d)}&\equiv i\int \!\frac{d^4l}{(2\pi)^4}\frac{1}{[l^2-m_1^2+i\epsilon][(p-l)^2-m_2^2+i\epsilon][(p-q_2-l)^2-m_3^2+i\epsilon][(l-q_1)^2-m_4^2+i\epsilon]}\nonumber\\
&\simeq\frac{-i}{16m_1m_2m_3m_4}\int \! \frac{d^4l}{(2\pi)^4}
\frac{1}{\left[l^0-\frac{\vec l^2}{2m_1}-m_1+i\epsilon\right]\left[l^0-M+\frac{\vec l^2}{2m_2}+m_2-i\epsilon\right]}\nonumber\\
&\times \frac{1}{\left[l^0+q_2^0-M+\frac{(\vec l+\vec q_2)^2}{2m_3}+m_3-i\epsilon\right]\left[l^0-q_1^0-\frac{(\vec l-\vec q_1)^2}{2m_4}-m_4+i\epsilon\right]}.
\end{align}
Here the possibility for two different cuts to go on-shell leads to a slightly more complicated three-dimensional integral
\begin{align}
-\frac{\mu_{12}\mu_{34}}{2m_1m_2m_3m_4}\int \! \frac{d^3l}{(2\pi)^3}\frac{1}{
[\vec l^2+d_{12}-i\epsilon][\vec l^2-2\frac{\mu_{34}}{m_4}\vec l\cdot \vec q_1-2\frac{\mu_{34}}{m_3}\vec l\cdot \vec q_2+d_{34}-i\epsilon]}\nonumber\\
\times\left[\frac{\mu_{24}}{[\vec l^2-2\frac{\mu_{24}}{m_4}\vec l\cdot \q_1+d_{24}-i\epsilon]}+\frac{\mu_{13}}{[\vec l^2+2\frac{\mu_{13}}{m_3}\vec l\cdot \q_2+d_{13}-i\epsilon]} \right], \label{eq:topo2}
\end{align}
where we defined
\begin{align}
 d_{12}&\equiv 2\mu_{12}\left(m_1+m_2-M\right), & d_{34}&\equiv 2\mu_{34}\left(m_3+m_4-q^0+\frac{\q_1^2}{2m_4}+\frac{\q_2^2}{2m_3}\right), \nn\\
 d_{24}&\equiv 2\mu_{24}\left(m_2+m_4-M+q_1^0+\frac{\q_1^2}{2m_4}\right), & d_{13}&\equiv 2\mu_{13}\left(m_1+m_3-M+q_2^0+\frac{\q_2^2}{2m_3}\right).
\end{align}
In both cases the remaining three-dimensional momentum integration needs to be carried out numerically.
\subsection{Tensor reduction}
%
Since each of the interactions of an $\Upsilon$ with a pair of bottom mesons scales with the momentum of the latter we will have to deal with
\begin{equation}
\frac{-\mu_{12}\mu_{23}\mu_{24}}{2m_1m_2m_3m_4}\int \!
\frac{d^3l}{(2\pi)^3}\frac{f(l)}{ [\vec l^2+c_{12}-i\epsilon][\vec
l^2+2\frac{\mu_{23}}{m_3}\vec l\cdot \vec q+c_{23}-i\epsilon][\vec
l^2-2\frac{\mu_{24}}{m_4}\vec l\cdot \vec q_1+c_{24}-i\epsilon]},
\end{equation}
where $f(l)=\{1,\, l^i,\, l^il^j\}$ for the fundamental scalar,
vector, and tensor integrals, respectively. Using the momentum of the
final state $\Upsilon$, $\vec q$, and $\vec q_\perp=\vec q_1-\vec
q(\vec q\cdot \vec q_1)/\vec q^2$, a convenient parametrization
reads
\begin{align}
J^{(1)i}&=\frac{-\mu_{12}\mu_{23}\mu_{24}}{2m_1m_2m_3m_4}\int \! \frac{d^3l}{(2\pi)^3}\frac{l^i}{[\vec l^2+c_1-i\epsilon][\vec l^2-2\frac{\mu_{23}}{m_3}\vec l\cdot \vec q+c_2-i\epsilon][\vec l^2-2\frac{\mu_{24}}{m_4}\vec l\cdot \q_1+c_3-i\epsilon]} \nn\\
&\equiv q^iJ_1^{(1c)}+q_\perp^iJ_2^{(1c)}
\end{align}
and
\begin{align}
J^{(2)ij}&=\frac{-\mu_{12}\mu_{23}\mu_{24}}{2m_1m_2m_3m_4}\int \! \frac{d^3l}{(2\pi)^3}\frac{l^il^j}{[\vec l^2+c_1-i\epsilon][\vec l^2-2\frac{\mu_{23}}{m_3}\vec l\cdot \vec q+c_2-i\epsilon][\vec l^2-2\frac{\mu_{24}}{m_4}\vec l\cdot \q_1+c_3-i\epsilon]} \nn\\
&\equiv\left(\delta^{ij}-\frac{q^iq^j}{\vec q^2}-\frac{q_\perp^iq_\perp^j}{\qperp^2}\right)J_0^{(2c)}
  +\frac{q^iq^j}{\vec q^2}J_1^{(2c)}
  +\frac{q_\perp^iq_\perp^j}{\qperp^2}J_2^{(2c)}
  +\frac{q^iq_\perp^j+q_\perp^iq^j}{|\q||\qperp|}J_3^{(2c)},
\end{align}
where the scalar integrals $J^{(r)}_m$ can easily be disentangled and have to be evaluated numerically. The corresponding expressions for topology II can be obtained by changing the denominators accordingly.
\subsection{Amplitudes}
\begin{table*}[t]
\begin{center}
\renewcommand{\arraystretch}{1.3}
\begin{tabular}{l|l}\hline\hline
Intermediate mesons &  Amplitude  \\ \hline
$[P,P,P,V]$  & $8 q_1\cdot q_2 J^{(2)ab}$ \\
$[P,V,P,V]$  & $-8 q_1\cdot q_2 J^{(2)ab}$  \\
$[V,P,V,P]$  & $4 \delta ^{ab} q_1^i \left(2 q_2^j J^{(2)ij}+q\cdot q_2 J^{(1)i}\right)$  \\
$[P,V,V,V]$  & $4 \delta ^{ab} \left(q_1^i q\cdot q_2-q_2^i q\cdot q_1\right) J^{(1)i}$  \\
$[V,P,V,V]$  & $4 \delta ^{ab} q_1^i \left(2 q_2^j J^{(2)ij}+q\cdot q_2 J^{(1)i}\right)$  \\
$[V,V,P,V]$  & $4 \delta ^{ab} \left(q_1^i q\cdot q_2-q_2^i q\cdot q_1\right) J^{(1)i}$  \\
$[V,V,V,P]$  & $8 \delta ^{ai} \delta ^{bj} q_1\cdot q_2 J^{(2)ij}+4 \delta ^{ab} q_1^i \left(2 q_2^j J^{(2)ij}+q\cdot q_2 J^{(1)i}\right)$  \\
$[V,V,V,V]$  & $4 \delta ^{ab} \left(4 \delta ^{ij} q_1\cdot q_2 J^{(2)ij}-q_2^i q\cdot q_1 J^{(1)i}+q_1^i \left(q\cdot q_2 J^{(1)i}-4 q_2^j J^{(2)ij}\right)\right)$  \\
\hline\hline%
\end{tabular}
\caption{\label{tab:amplitudes_c1} All loops contributing to topology (c1) in Fig.~\ref{fig.FeynmanDiagram}. The mesons are listed as $[M1,M2,M3,M4]$, $P$ and $V$ denote intermediate pseudoscalar and vector mesons, respectively. The different flavors are dropped for simplicity---the full amplitude contains the sum of all possible ones.}
\end{center}
\end{table*}

Tables~\ref{tab:amplitudes_c1} and \ref{tab:amplitudes_d1}  list the relevant amplitudes for this calculation.
We will only give the dominant amplitudes, \textit{i.e.}\ the ones that contribute to the part proportional to $\epsilon(\Upsilon')\cdot\epsilon(\Upsilon)$ as was explained in the main text.
We further notice that all box diagrams are proportional to the overall factor $\epsilon^a(\Upsilon')\,\epsilon^b(\Upsilon)\,g_\pi^2\,g_{J'HH}\,g_{JHH}/F_\pi^2$.

Finally, we need to consider the different flavors of the intermediate bottom mesons. For topology (c1) with a pair of charged pions four possibilities exist:
$[B^{(*)+},B^{(*)-},B^{(*)+},B^{(*)0}]$, $[B^{(*)-},B^{(*)+},B^{(*)-},{\bar B}^{(*)0}]$, $[B^{(*)0},{\bar B}^{(*)0},B^{(*)0},B^{(*)+}]$, and  $[{\bar B}^{(*)0},B^{(*)0},{\bar B}^{(*)0},B^{(*)-}]$. For topology (d1) this reduces to just two:
$[B^{(*)+},B^{(*)-},{\bar B}^{(*)0},B^{(*)0}]$ and $[{\bar B}^{(*)0},B^{(*)0},B^{(*)+},{\bar B}^{(*)-}]$. For the case of neutral pions the number of possible diagrams doubles---a factor 2 that is balanced by the factor $\sqrt2$ in the $SU(3)$ light-meson matrix.

\begin{table*}[t]
\begin{center}
\renewcommand{\arraystretch}{1.3}
\begin{tabular}{l|l}\hline\hline
Intermediate mesons &  Amplitude  \\ \hline
$[P,P,V,V]$  & $8 q_1\cdot q_2 J^{(2)ab}$ \\
$[P,V,P,V]$  & $4\delta ^{ab} \left(q_1\cdot q_2 \left(2 \delta ^{ij} J^{(2)ij}+q_2^i J^{(1)i}\right)- q_1^i \left(2 q_2^j J^{(2)ij}+q_2{}^2 J^{(1)i}\right)\right)-8 \delta ^{ai} \delta ^{bj} q_1\cdot q_2 J^{(2)ij}$  \\
$[V,V,P,P]$  & $4 \delta ^{ab} q_1^i \left(2 q_2^j J^{(2)ij}+q\cdot q_2 J^{(1)i}\right)$ \\
$[V,P,V,P]$  & $4 \delta ^{ab} q_1^i \left(\left(q\cdot q_2+2 q_1\cdot q_2\right) J^{(1)i}-2 q_2^j J^{(2)ij}\right)$ \\
$[P,V,V,V]$  & $-4 \delta ^{ab} \left(2 q_1^i q_2^j J^{(2)ij}+q_2^i \left(q_1\cdot q_2-q_1{}^2\right) J^{(1)i}\right)$ \\
$[V,P,V,V]$  & $-4 \delta ^{ab} q_1^i \left(2 q_2^j J^{(2)ij}-\left(q\cdot q_2+2 q_1\cdot q_2\right) J^{(1)i}\right)$ \\
$[V,V,P,V]$  & $-4 \delta ^{ab} \left(2 q_1^i q_2^j J^{(2)ij}+q_2^i \left(q_1\cdot q_2-q_1{}^2\right) J^{(1)i}\right)$ \\
$[V,V,V,P]$  & $-4 \delta ^{ab} q_1^i \left(2 q_2^j J^{(2)ij}+\left(q_2{}^2-q_1\cdot q_2\right) J^{(1)i}\right)$ \\
$[V,V,V,V]$  & $4 \delta ^{ab} \left(4 \delta ^{ij} q_1\cdot q_2 J^{(2)ij}-q_2^i q\cdot q_1 J^{(1)i}+q_1^i \left(q\cdot q_2 J^{(1)i}-4 q_2^j J^{(2)ij}\right)\right)$ \\
\hline\hline%
\end{tabular}
\caption{\label{tab:amplitudes_d1} All loops contributing to topology (d1) in Fig.~\ref{fig.FeynmanDiagram}; see Table~\ref{tab:amplitudes_c1} for further notation.}
\end{center}
\end{table*}

\end{appendix}

\end{document}